\documentclass{nature}
\usepackage{graphicx}

\title{Enrichment by supernovae in  globular clusters with multiple populations}

\author{Jae-Woo Lee$^{1}$,
Young-Woon Kang$^{1}$,
Jina Lee$^{1}$ \&
Young-Wook Lee$^{2}$
}

\begin{document}
\spacing{1}

\maketitle

\begin{affiliations}
\item Department of Astronomy and Space Science, ARCSEC,
Sejong University, Seoul 143-747, Korea
\item Center for Space Astrophysics, Yonsei University, Seoul 120-749,
  Korea
\end{affiliations}

\begin{abstract}
The most massive globular cluster in the Milky Way, $\omega$ Centauri, is
thought to be the remaining core of a disrupted dwarf galaxy\cite{lee99,bf03},
as expected within the model of hierarchical merging\cite{freeman,diemand}.
It contains several stellar populations having different heavy elemental abundances
supplied by supernovae\cite{johnson} --- a process known as metal
enrichment. Although M22 appears to be similar to $\omega$  Cen\cite{marino},
other peculiar globular clusters do not\cite{carretta,georgiev}.
Therefore $\omega$  Cen and M22 are
viewed as exceptional, and the presence of chemical inhomogeneities
in other clusters is seen as `pollution' from the intermediate-mass
asymptotic-giant-branch stars expected in normal globular
clusters\cite{ventura}.
Here we report Ca abundances for seven globular clusters
and compare them to $\omega$ Cen. Calcium and other heavy elements can
only be supplied through numerous supernovae explosions of
massive stars in these stellar systems\cite{timmes},
but the gravitational potentials of the present-day clusters
cannot preserve most of the ejecta from such explosions\cite{baum}.
We conclude that these globular clusters,
like $\omega$  Cen, are most probably the relics of more massive primeval
dwarf galaxies that merged and disrupted to form the proto-Galaxy.
\end{abstract}

The Sejong/ARCSEC Ca uvby survey program was initiated in 2006
to investigate the homogenous metallicity scale for globular clusters
and to obtain the complete metallicity distribution function
of the Galactic bulge using the $hk$ index [= $(Ca - b) - (b - y)$]\cite{att}.
The Ca filter in the $hk$ index measures ionized calcium H and K lines,
which have been frequently used to calibrate metallicity scale
for globular clusters\cite{zinn80, zw}.
The utility of the $hk$ index is that it is known to be about three times
more sensitive to metallicity than the $m_1$ index is for stars
more metal-poor than the Sun and it has half the sensitivity
of the $m_1$ index to interstellar reddening\cite{att}.
During the last three years, we have used more than 85 nights of
CTIO 1.0-m telescope time for this project.
The telescope was equipped with an STA 4k $\times$ 4k CCD camera,
providing a plate scale of 0.289 arcsec/pixel and
a field of view of 20 $\times$ 20 arcmin.
All of our targets accompanied with standards were observed
under the photometric weather conditions and most of targets were
repeatedly visited between separate runs.
The photometry of our targets and standards were analyzed using
DAOPHOT II, ALLSTAR, and ALLFRAME\cite{daophot, allframe}.

In the course of metallicity calibration of red giant branch (RGB)
stars in GCs, we found that many GCs show split in the RGB
in their $hk$ versus $V$ color-magnitude diagrams (Figs 1 and 2).
The prime examples are M22 and NGC1851.
In particular, the double RGB sequence in M22 is very intriguing.
The differential reddening effect and the contamination from
the off cluster populations cannot explain the double RGB sequences in M22
(see Supplementary Information).
It has been debated for decades whether this cluster is
chemically inhomogeneous or not, but
the recent high resolution spectroscopic study of 17 RGB stars in the cluster
suggests that it contains chemically inhomogeneous subpopulations\cite{marino}.
The bimodality in the $m_1$ index of M22 RGB stars was also known,
but it has been argued that it is most likely due to
the bimodal CN abundances, where CN absorption strengths
strongly affect the $m_1$ index, not due to the bimodal distribution of
heavy elements in the cluster\cite{norris, richter, attm20}.
The star-to-star light elemental abundance (C, N, O, Na, Mg and Al)
variations have been known for decades and
they are now generally believed to be resulted from chemical pollutions
by intermediate-mass asymptotic giant branch stars\cite{ventura}
or fast rotating massive stars\cite{decressin}.
However, it should be emphasized that our $hk$ measurements for RGB stars
in M22, NGC1851 and other GCs show discrete or bimodal distributions
in calcium abundance, which cannot be supplied by intermediate-mass
asymptotic giant branch stars or fast rotating massive stars.

As shown in Fig 3, the difference in calcium, silicon, titanium
and iron abundances between the calcium weak (Ca-w hereafter) group
with smaller $hk$ index and the calcium strong (Ca-s hereafter) group
with larger $hk$ index in M22 and NGC1851 suggests that
they are indeed chemically distinct\cite{bw, n1851, yong08, cudworth}.
(It is not shown in the figure but europium also has a bimodal abundance
distribution in M22, in the sense that the Ca-s group has
a higher europium abundance.)
As for the origin of chemical inhomogeneity in globular clusters,
at least four viable chemical enrichment mechanisms have been proposed up to date.
They are, in the order of time required to emerge;
(i) fast rotating massive stars, (ii) Type II supernovae, 
(iii) intermediate-mass asymptotic giant branch stars, and (iv) Type Ia supernovae.
If the current understanding of supernovae explosions is correct,
only Type Ia and II supernovae can supply the heavy elements
such as calcium and iron\cite{timmes}.
To explain the discrete calcium abundances seen in M22 and NGC1851, however,
the contribution from Type Ia supernovae can be ruled out for two reasons.
First, the longer timescale ($\geq$ 1 -- 2 Gyr) before the onset of
Type Ia supernova explosions, which would produce detectable age spread
between two populations; and second,
the enhanced $\alpha$-elemental abundances, indicative of absence of
contributions from Type Ia supernovae\cite{timmes}.
Qualitatively, the differences in elemental abundances
between the two stellar populations in M22 and NGC1851 can be naturally
explained by invoking chemical enrichment by Type II supernovae,
where $\alpha$-elements (silicon, calcium, and titanium) and
$r$-process element (europium) are dominantly produced.
However, our results do not necessarily imply that Type II supernovae
are solely responsible for the chemical enrichment in M22 and NGC1851,
since all four above-mentioned mechanisms may be involved.
We emphasize that the crux of our results is the undeniable evidence for
Type II supernovae contribution to chemical enrichment of some globular clusters,
in sharp contrast to the widely accepted idea of chemical pollution
only by intermediate-mass asymptotic giant branch or fast rotating massive stars,
with which the chemical enhancement of the $\alpha$-
and $r$-process elements in the second generation of the stars
cannot be easily explained.

More than half of 37 globular clusters in our sample shows discrete or
broad distributions of the $hk$ index in their RGB sequences.
In Fig 2, We show color-magnitude diagrams for some of exemplary globular clusters
in the order of $hk$ widths of RGB sequences at $V_{HB}$, the $V$ magnitude level
at the horizontal branch:
$\omega$ Cen, M22, NGC1851, NGC2808, M4, M5, NGC6752
and NGC6397 (see also Supplementary Table 3 and Figs 6 -- 13).
NGC2808 is known to have multiple main-sequences but
no multiple RGB sequences have been reported to date.
Our new results show that NGC2808 shows at least two discrete
RGB sequences with a large spread in calcium abundance.
Similarly, M5 has very broad $hk$ index in the RGB sequence
and NGC6752 shows discrete RGB sequences.
It is interesting to note that all the globular clusters with signs of multiple
stellar populations have relatively extended horizontal branch,
while the globular clusters with normal horizontal branch
(e.g. NGC6397 in Fig 2 and Supplementary Fig 13) show
no spread or split in RGB.
This is consistent with the suggestion that the extended horizontal branch is
a signal of the presence of multiple stellar populations in globular clusters\cite{lee07}.

The overwhelming problem of the chemical enrichment by Type II supernovae
in globular clusters is that their ejecta are considered to be too energetic
to be retained by less massive systems like typical Galactic globular clusters
($\leq$ a few times 10$^5$ $M_\odot$)\cite{baum}.
Our results therefore suggest that M22, NGC1851 and other globular clusters with RGB split
were much more massive in the past,
unless the current understanding of supernovae explosions is in great error.
Perhaps, these globular clusters were once nuclei of dwarf-galaxy-like fragments and
then accreted and dissolved in the Milky Way, as is widely accepted for
$\omega$ Cen\cite{lee99, bf03, piotto05}.
Recent calculations suggest that a massive
($\geq$ a few times $10^6$ $M_\odot$) star cluster embedded
in a proto-dwarf galaxy could accrete gas from its host dwarf galaxy
which may cause the formation of the second generation stars,
producing multiple stellar populations\cite{pflamm}.
Note that this scenario is also suggesting that the globular clusters with multiple
stellar populations would be the remaining cores of the proto-galactic
building blocks.
This idea is supported by the recent investigations of the extended horizontal branch globular clusters
(i.e. globular clusters with signatures of multiple stellar populations),
which has shown that extended horizontal branch globular clusters are clearly distinct from
the normal globular clusters in orbital kinematics and mass\cite{lee07}.
Extensive photometric surveys for fainter stars in these globular clusters,
as well as spectroscopic surveys for stars in double RGB sequences,
would undoubtedly help to shed more light into the discovery reported here.

\begin{addendum}
\item[Supplementary Information]is linked to the online version of the paper
at www.nature.com/nature.
 \item J.-W. L. thanks A. Walker for providing the CTIO Ca filter
transmission curve, D. Yong for NGC1851 spectroscopic data before publication
and A. Yushchenko for discussions on spectrum synthesis.
Support for this work was provided by the National Research
Foundation of Korea to the Astrophysical Research Center for the Structure and
Evolution of the Cosmos (ARCSEC).
This work was based on observations made with the CTIO 1.0-m telescope,
which is operated by the SMARTS consortium.
\item[Author Contributions] J.-W. L. performed observations,
data analysis, interpretation, model simulations and writing of
the manuscript.
Y.-W. K. participated in observation planning,
J. L. performed a part of observations and data analysis.
Y.-W. L. performed interpretation and writing of the manuscript.
All authors discussed the results and commented on the manuscript.
 \item[Author Information] Reprints and permissions information is
available at www.nature.com/reprints.
The authors declare that they have no competing financial interests.
Correspondence should be addressed to J.-W. L. (jaewoolee@sejong.ac.kr) or
Y.-W. L. (ywlee2@yonsei.ac.kr).
\end{addendum}

\clearpage

\begin{figure*}
  \begin{center}
  \includegraphics[width=1.\textwidth]{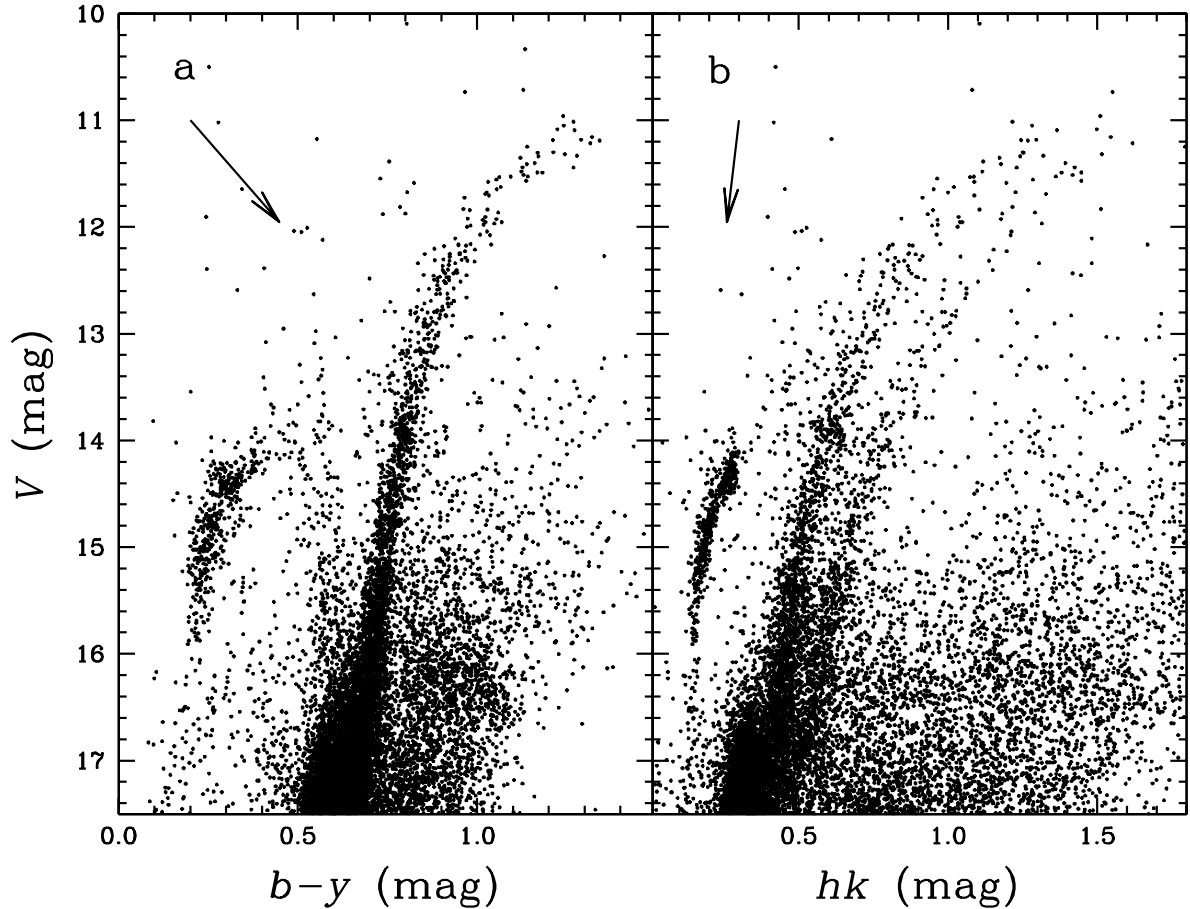}
  \end{center}
\caption{{\bf Color-magnitude diagrams for M22.}
{\bf a,} $V$ versus $b-y$; {\bf b,} $V$ versus $hk$.
In {\bf b}, note the distinct and discrete double RGB sequences in M22.
This cannot be due to differential reddening effect across the cluster
or the contamination from the off cluster field but, is due to the difference
in calcium abundance, which was synthesized in supernovae,
between the two RGB sequences.
The number ratio between the Ca-w group with smaller $hk$ index and
the Ca-s group with larger $hk$ index is about 70:30.
Black arrows in each panel denote reddening vectors.
 \label{fig1}}
\end{figure*}

\begin{figure*}
  \begin{center}
  \includegraphics[width=1.\textwidth]{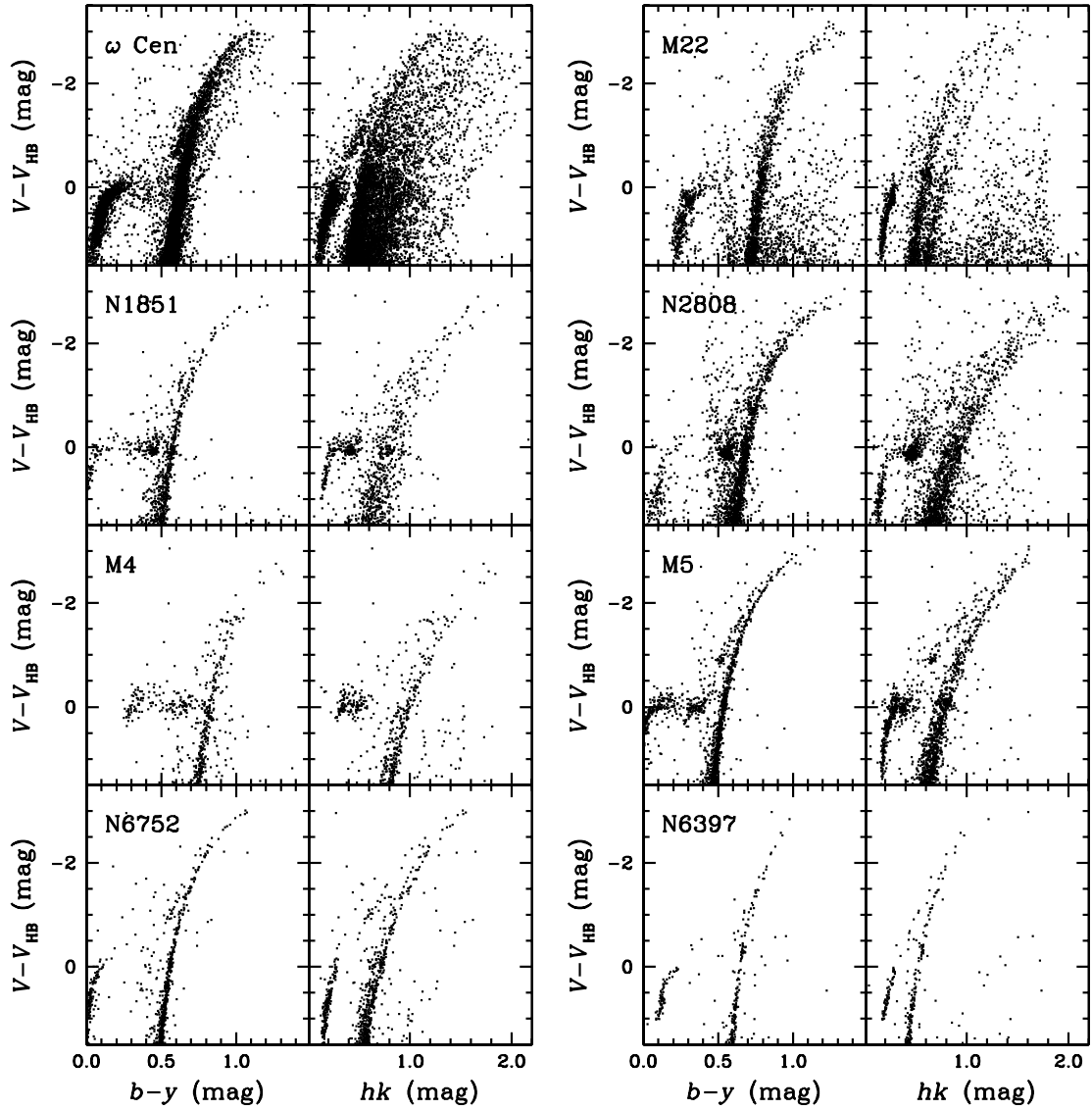}
  \end{center}
\caption{{\bf Color-magnitude diagrams for $\omega$ Cen, M22, NGC1851, NGC2808,
M4, M5, NGC6752 and NGC6397.}
Note that, while the distributions
of the RGB sequences in the $b-y$ color are relatively narrow,
those in the $hk$ index are either discrete or broad.
This is evidence for the multiple stellar populations with
distinct calcium abundances.
Among these globular clusters, NGC6397 appears to be the only normal globular
cluster with simple population (i.e. coeval and monometallic).
 \label{fig2}}
\end{figure*}

\begin{figure*}
  \begin{center}
  \includegraphics[width=.85\textwidth]{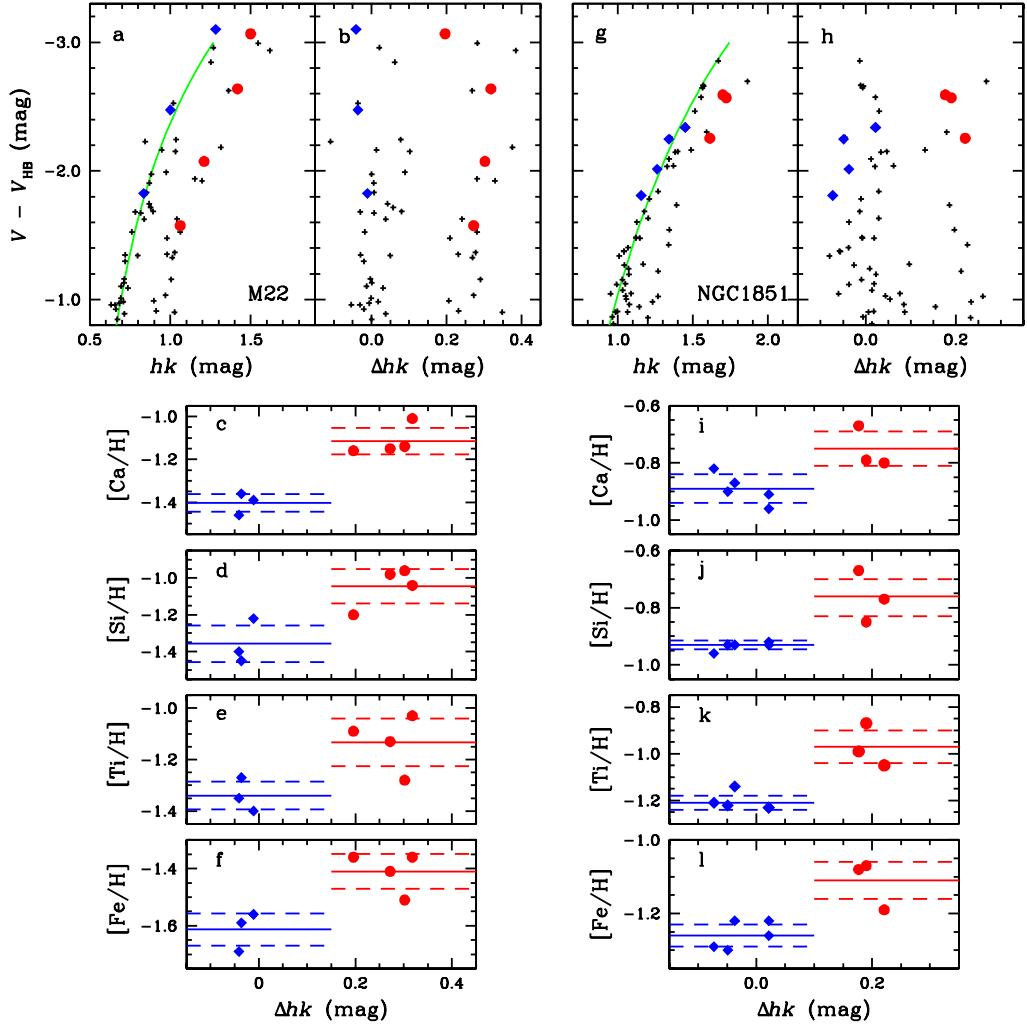}
  \end{center}
\caption{{\bf Differences in chemical compositions between double
RGB sequences in M22 and NGC1851.}
{\bf a, b,} Black `plus' signs denote stars in M22 with proper motion
membership probabilities $P$ $\geq$ 90\%;  blue filled diamonds
and red filled circles denote RGB stars studied with
high-resolution spectroscopy in the Ca-w and the Ca-s groups,
respectively\cite{cudworth, bw}.
The green solid line denotes the fiducial sequence of RGB stars
and $\Delta hk$ denotes the difference in the $hk$ index against
the fiducial sequence.
The double RGB sequences persist in proper motion member stars.
{\bf c -- f,} Comparisons of elemental abundances between
the Ca-w and the Ca-s groups in M22.
Solid lines denote the mean values, and dashed lines denote standard deviations
of each group. The Ca-s group has higher $\alpha$-elements (Si, Ca, and Ti)
and iron abundances, which must be supplied by numerous
Type II supernova explosions.
{\bf g, h,} Black `plus' signs denote stars in NGC1851; blue filled diamonds
and red filled circles denote RGB stars studied with high-resolution
spectroscopy in the Ca-w and the Ca-s groups,
respectively\cite{yong08, n1851}.
{\bf i -- l,} As {\bf c -- f} but for NGC1851.
 \label{fig3}}
\end{figure*}

\clearpage

\noindent
{\bf {\Large Supplementary Information}} 
\normalsize

\section{The CTIO Ca Filter System}
\subsection{The Filter Transmission}
The Ca filter system was designed to include Ca II H and K lines at
$\lambda$ 3968 and 3933 \AA, respectively, with a full-width half maximum (FWHM)
of approximately 90 \AA.
The CN band absorption strengths at $\approx$ $\lambda$ 3885 \AA~ are
often very strong  in stellar spectra and the lower limit of the Ca filter
is set to avoid  contamination by the CN band$^{12}$.
The Ca filter used at Cerro Tololo Inter-American Observatory (CTIO)
has a similar FWHM, approximately 90 \AA, but its passband is shifted
approximately 15 \AA~ to the longer wavelength 
(Alistair Walker, private communication) compared to that in
Anthony-Twarog \emph{et al.}$^{12}$
In Supplementary Figure \ref{sifig:filter}, we show the transmission
functions of the both Ca filters.
In the figure, we also show synthetic spectra for the CN normal and
the CN strong RGB stars in typical intermediate metallicity globular clusters
(GCs) as an illustration.
The effect of the CN band on the $hk$ index is
negligible as will be discussed below.

\subsection{The Central Wavelength Drift of the Ca Filter}
The CTIO Ca filter transmission function shown
in Supplementary Figure \ref{sifig:filter} is
that measured with a collimated beam.
It is known that the passband of the narrow band interference filter
depends on the angle of the incidence beam following,
\begin{equation}
 \lambda = \lambda_0\left( 1-\frac{\sin^{2}\beta}{n^{*2}}\right) ^{1/2},
\end{equation}
where $\lambda_0$ is the wavelength of peak transmittance at normal incidence,
$\beta$ is the angle of incidence of the collimated beam on the filter and 
$n^*$ is the effective refractive index of the filter\cite{clarke}.
Therefore, when the Ca filter is used with a fast telescope,
the filter passband can be significantly different from that shown
in Supplementary Figure \ref{sifig:filter}.
The CTIO 1-m telescope used for our survey
is a slow telescope with $f$/10.5 and the effect resulted from
the angular dependency of a converging beam is expected to be very small.
Assuming $\beta$ $\approx$ 1/21 radians for the converging beam
at the CTIO 1-m telescope and $n^*$ $\approx$ 1.4 for the CTIO Ca filter,
the peak wavelength of the Ca filter will be shifted by 2.3 \AA~
to the shorter wavelength.
Given the much larger FWHM of the CTIO Ca filter,
this may contribute small effect.
We investigate contributions to the $hk$ index resulted from
the shifted Ca passband using synthetic spectra for
typical intermediate metallicity RGB stars in our GCs.
Our calculations integrating over the filter transmission curve
show that this effect contributes no larger than 0.011 mag
to our $hk$ measurements, in the sense that the shifted Ca passband
to the shorter wavelength gives slightly larger $hk$ values.
We emphasize that, since our results are based on a single instrument setup
(the same telescope, filters and the CCD camera)
during the observations of our science targets and the photometric standards,
this effect is expected to be cancelled out during our photometric calibrations.
Also importantly, our main results presented here rely on the split
or the spread in the $hk$ index of RGB stars of an individual GC.
Therefore, the shifted Ca passband affects similar degree to
the $hk$ index of the RGB stars in a GC and does not contribute
to the apparent RGB split or the spread in the $hk$ index of
an individual GC.

\subsection{Effect of radial motions and internal velocity dispersions of GCs}
The mean radial motion of 139 GCs in our Milky Way Galaxy\cite{harris} is
$|v_r|$ = 110 km/s, equivalent to the wavelength shift by
$|\Delta\lambda|$ $\approx$ 1.4 \AA~ at $\lambda$ 3950\AA.
We calculate the contribution due to the mean radial motion of GCs
to the $hk$ index using the shifted CTIO Ca passband and synthetic spectra
for typical intermediate metallicity GC RGB stars.
We find that the net effect is negligibly small, $|\Delta hk|$ $<$ 0.006 mag.
Among our eight GCs, NGC1851 has the largest radial velocity, $v_r$ = 321 km/s,
equivalent to the wavelength shift by $\Delta \lambda$ = 4.2 \AA~ to the
longer wavelength at $\lambda$ 3950\AA.
We calculate the contribution due to the radial motion of NGC1851
using the shifted CTIO Ca passband and the red-shifted synthetic spectra
with a fixed CN abundance for the cluster.
We obtain $\Delta hk$ $\approx$ 0.015 mag, in the sense that
the red-shift is resulted in a slightly larger $hk$ index.
As we discussed above, the difference in the $hk$ index due to the high radial
motion of NGC1851 does not affect our results presented here,
since the $hk$ indices of RGB stars in NGC1851 will be affected by similar degree
and the mean radial motion of the cluster does not produce
an apparent split or a spread in the $hk$ index.
Perhaps, this effect may become important in the inter-cluster comparisons,
which is beyond the scope of our study.

What concerns us most about the high radial velocities of some GCs,
in particular for red-shift, is the potential contamination
by the strong CN band absorption features at $\lambda$ 3885 \AA~
as shown in Supplementary Figure \ref{sifig:filter}.
For example, NGC1851 has a bimodal CN distribution and some RGB stars
show very strong CN band absorption strengths\cite{hesser}.
Due to its high radial velocity away from us (i.e. red-shifted),
the CN band absorption features in the CN-strong RGB stars
may affect the $hk$ index and, subsequently, may produce
an apparent RGB split of the cluster as shown in Figure 2 or
Supplementary Figure \ref{sifig:n1851dist}.
We calculate the CN band contributions
using the shifted CTIO Ca passband and the red-shifted synthetic spectra for
the CN-normal and the CN-strong RGB stars (see discussion below).
Our calculations suggest that the net effect is negligibly small,
$\Delta hk$ $\leq$ 0.003 mag, and the high radial velocity
of NGC1851 combined with a bimodal CN distributions
does not produce the RGB split in the $hk$ index.

We also investigate the effect of the internal velocity dispersion
of an individual GC. Assuming $\sigma_{LOS}$  = 15 km/s,
equivalent to $\Delta\lambda$ $\leq$ 0.2 \AA~ at $\lambda$ 3950\AA,
we obtain $\Delta hk$ $\leq$ 0.001 mag following the same method 
described above, and the effect from the internal velocity dispersions of GCs
does not affect our results.

\subsection{Summary of uncertainties on the $hk$ index}
Supplementary Table \ref{tab:exterr} summarizes the uncertainties
in our $hk$ index measurements relevant to the CTIO Ca passband.
(The variations in the $hk$ index due to differences in elemental abundances
of GC RGB stars will be discussed below.)
As discussed above, the effects due to the shifted CTIO Ca passband
and the radial motions of GCs do not affect our results,
since both effects contribute similar degree to the $hk$ indices
among RGB stars in a GC.
(i.e. They only affect the zero point of the $hk$ index
and they do not affect the $\Delta hk$ distributions).
In addition to our photometric measurement errors which will be discussed below,
the effects due to the internal velocity dispersions of GCs
and the differential interstellar Ca II absorption (see discussion below)
can affect our $hk$ index measurements.
However, their contributions to our $hk$ measurements are no larger
than 0.022 mag and they do not affect our main conclusion
presented here.
Therefore, our results strongly suggest that the split or the spread
in the $hk$ index of RGB stars in GCs are related to
the variations in elemental, in particular calcium,
abundances among RGB stars in a GC, which will be discussed below.

\section{The Double RGB Sequences of M22}

\subsection{Differential Reddening Effect on the Double RGB Sequences in M22}
The continuous interstellar extinction by the interstellar dust
and the discrete interstellar line extinction by the interstellar Ca II atoms
may affect our main results.
We considered both effects and will discuss that the RGB split of M22
in the $hk$ index is indeed due to the difference in calcium abundances
between two stellar populations in M22 and
other explanations are highly unlikely.
Also both effects tend to produce spreads in RGB sequences rather than
the distinct and discrete RGB sequences of GCs reported here.

The differential continuous reddening across the cluster
can thicken the apparent RGB sequence of GCs
in broad-band optical photometry\cite{monaco}.
In contrast to other color indices being used in broad-band photometry,
the $hk$ index is known to be insensitive to interstellar reddening$^{12}$,
$E(hk)/E(b-y)$ = $-$0.16 and $E(hk)/E(B-V)$ = $-$0.12.
The difference in the $hk$ index between the two RGB sequences in M22
is about 0.2 mag at the magnitude level of the horizontal branch.
If this $hk$ split is only due to differential reddening effect,
we would expect even larger separation of the two RGB sequences
in the $b-y$ color and the $V$ magnitude.
The reddening correction value in the $b-y$ color, $E(b-y)$,
for the Ca-s group is about $-$1.25 mag,
equivalent to $E(B-V)$ = $-$1.69 mag
assuming $E(b-y)/E(B-V)$ = 0.74, making the RGB stars in the Ca-s group
too hot to be RGB stars (see Supplementary Figure \ref{sifig:reddening} -- d).
At the same time, the extinction correction value in the $V$ magnitude is
$-$5.24 mag, assuming $A_V$ = 3.1$\times E(B-V)$, for the Ca-s RGB stars.
Applying this large extinction correction makes the RGB stars
in the Ca-s group too bright to be members of M22
(see Supplementary Figure \ref{sifig:reddening} -- e \& f).
We emphasize that both the Ca-w and the Ca-s groups are proper motion
members of the cluster as shown in Figure 3.
Note also that the reddening vector
(see Figure~1 or Supplementary Figure \ref{sifig:m22cmd}) is
almost parallel to the slopes of HB and RGB in the $hk$ versus $V$ CMD,
and thus the differential reddening can not produce the RGB split.
Therefore, continuous differential reddening effect can be completely
ruled out to explain the observed bimodal RGB sequences in M22.
Similarly, the interstellar reddening toward NGC1851
is very small, $E(B-V)$ = 0.02 mag\cite{harris}, but the $hk$ split in RGB stars
of the cluster is as large as 0.2 mag
(see Supplementary Figure \ref{sifig:n1851dist}), which can not be explained
by differential continuous reddening effect$^{22}$.

The previous study for the GCs
showed that the equivalent width of the interstellar Ca II K absorption
line strength can be as large as several times 100 m\AA\cite{beers}.
The interstellar Ca II atom is thought to be heavily depleted on to dust
in denser clouds\cite{hunter}, which may cause small-scale 
differential discrete reddening effect across M22 and other GCs studied here
(see also Andrew \emph{et al.}\cite{m92} for the small-scale variations
of interstellar Na I D lines\footnote{Note that the number of interstellar Na I
atoms appears to be about a factor of ten larger than that of
interstellar Ca II atoms\cite{hunter}.}
 toward the less extincted globular cluster M92).
We generate synthetic spectrum to surrogate interstellar Ca II H \& K
absorption lines. We adopt a gaussian line profile with a FWHM of 1 \AA,
equivalent to $\Delta v_r$ $\approx$ 76 km/s, and we assign equivalent widths
of 350 m\AA~ and 650 m\AA~ for the interstellar Ca II H \& K lines,
respectively.
Our synthetic spectrum is shown in Supplementary Figure \ref{sifig:filter} -- (c).
Assuming they are linear part of the curve of growth\cite{smoker},
the column density of the interstellar Ca II can be estimated as
\begin{equation}
 N({\rm Ca~II}) = 1.13\times10^{20}\frac{EW}{\lambda^2f},
\end{equation}
where $EW$ and $\lambda$ are the equivalent width and wavelength in \AA~ and
$f$ is the oscillator strength.
Using the oscillator strengths of 0.681 and 0.341 for
Ca II H \& K, respectively, the column density for
interstellar Ca II is $\log N$(Ca II) $\approx$ 12.8 cm$^{-2}$,
equivalent to $\Delta E(B-V)$ $\approx$ 0.32 mag\cite{hunter}.
If this large amount of small-scale interstellar Ca II variation exists
among our GCs studied here, how much will it affect our $hk$ index measurements?
We calculate the $Ca$ magnitudes with and without
the interstellar Ca II variations using the shifted CTIO Ca transmission function.
The difference in the $Ca$ magnitude (i.e. in the $hk$ index since
the interstellar Ca II H \& K lines do not affect $b$ or $y$ passbands)
is only 0.010 mag and, therefore, the differential discrete
reddening effect due to the variations in the interstellar Ca II abundances
can be completely ruled out to explain the GC RGB splits in the $hk$ index.

\subsection{The Spatial Distributions}
In Supplementary Figure \ref{sifig:spatial}, we show the spatial distributions
of Ca-w and Ca-s RGB stars in M22.
As can be seen in the figure, each population
does not show any spatially patched features, supporting our results that
differential reddening is not responsible for the RGB split in M22.

\subsection{Contamination from the Milky Way's Bulge Population}
M22 is located in the direction of the Milky Way's bulge and
the contamination from the bulge population may affect our results.
However, this is very unlikely, since the proper motion member RGB stars show
discrete double RGB sequences as shown in Figure 3.
In addition, the bulge RGB stars are located farther from the Sun,
more metal-rich and suffering from heavier interstellar reddening than
those in M22 are. In Supplementary Figure \ref{sifig:m22cmd},
we compare M22 CMDs with those of two bulge fields (NGC6528 and OGLEII - 12).
As can be seen in the figure, the RGB stars in the bulge are fainter and redder
than those in M22 are and the contamination from the Milky Way's
bulge population does not affect our results.

\subsection{Effects of Metal Contents and Helium Abundances on the M22 RGB}
As shown in Figure 3, the stars in the Ca-s group are about 0.2 dex
more metal-rich than those in the Ca-w group.
It is suspected that this large metallicity spread may produce
any detectable discrepancy in stellar evolutionary sequences,
in particular for RGB sequence, between two stellar populations
based on broad-band photometry.
To explore metallicity effect on the RGB sequence, we compare $BV$ CMD
by Monaco \emph{et al.}\cite{monaco} with the latest $Y^2$ isochrones
(Version 3, Yi \emph{et al.} in preparation).
In Supplementary Figure \ref{sifig:iso}, we show model isochrones for
[Fe/H] = $-$1.6 and $-$1.4 with the helium abundance of $Y$ = 0.23 and
the age of 11 Gyr, using the reddening value and the distance modulus
for the cluster from Harris\cite{harris}.
Although the split in the RGB sequences of two model isochrones is noticeable,
the discrepancy in the RGB sequence does not appear to cause a serious problem
to explain the $BV$ CMD by Monaco \emph{et al.}\cite{monaco}
Note that the $V$ magnitude difference in the sub-giant branch between
two model isochrones can be as large as 0.2 mag, apparently consistent with
recent HST/ACS observations of the cluster\cite{piotto09}.

As inferred from the extended HB (EHB) morphology of M22,
the second generation of the stars
is expected to have enhanced helium abundance by
$\Delta Y \approx$ 0.05\cite{dantona02}.
Since the new version of $Y^2$ isochrones provides models with
enhanced helium abundances, we investigate the effect of helium abundance
on the evolutionary sequence.
As illustrated in Supplementary Figure \ref{sifig:iso} -- (c),
the discrepancy between two stellar populations alleviates due to the opposite
effect of metal contents and helium abundances on the RGB temperature.
Since the second generation of the stars in M22 shows signs of the chemical
enrichment by Type II supernovae and intermediate-mass asymptotic giant branch
(AGB) stars,
the second generation of stars may be slightly younger than the first generation.
Assuming the age difference of 1 Gyr between two stellar populations,
two model isochrones are in excellent agreement except for bright RGB sequence,
where the observed number of stars is small.

It is intriguing to note that the number ratio between the Ca-w and
the Ca-s RGB stars (70:30) found here is very similar to
those found between
(1) the two stellar groups with different [Fe/H] and [$s$-process/Fe]
ratios$^{6}$,
(2) the brighter SGB and the fainter SGB stars\cite{piotto09}, and
(3) the two groups of HB stars with the bluer HB being less populated.
The population synthesis models (Han \emph{et al.} 2009, in preparation)
suggest that this can be naturally reproduced by the enhanced metal and
helium abundances in the second generation of stars.

\section{The \emph{hk} and Metallicity Distributions of GCs}
\subsection{Observations}
In Supplementary Table~\ref{tab:log}, we show the journal of observations
for eight GCs. They were observed under the photometric weather conditions
and, for most cases, the median seeing was about
1.5 -- 1.6 arcsec during our observations.
Note that the RGB stars in NGC2808 are roughly 21 times fainter than those
in NGC6397 in the Ca passband for a fixed magnitude,
while our total integration time for NGC2808 is only about three times longer
than that for NGC6397 in the Ca passband.
Statistically, the lack of total integration time for NGC2808 will be resulted
in a $\approx$ 2.6 times larger $Ca$ measurement error than that expected for
NGC6397 at a given $Ca$ magnitude, particularly for fainter stars.
Although our survey relied on a rather small telescope through
a rather narrow filter at a rather blue wavelength\footnote{Fortunately,
the CCD camera used in our survey has rather high
quantum efficiency (QE) at shorter wavelength with QE $\approx$ 0.686 at 
$\lambda$ 3800\AA~ and  $\approx$ 0.770 at $\lambda$ 4000\AA.
(see http://www.astronomy.ohio-state.edu/Y4KCam/OSU4K/index.html\#DQE).},
our investigations of the multiple stellar populations of GCs are
focused on bright RGB stars, where the numbers of photon in the Ca passband
are enough so that the measurement errors,
including propagation errors during the photometric calibrations,
are less than 0.020 mag (see Supplementary Table~\ref{tab:error}).

\subsection{Color Distributions of Bright RGB Stars}
Here, we investigate the $b-y$ color and the $hk$ index distributions of
RGB stars brighter than $V-V_{\rm HB}$ = 1.0 mag.
We derive lower order ($\approx$ 4 -- 5) polynomial fits,
which are forced to pass through the peak $b-y$ colors or 
the peak $hk$ indices of
given magnitude bins, as fiducial sequences for eight GCs and then
we calculate differences in the $b-y$ color, $\Delta(b-y)$, 
or the $hk$ index, $\Delta hk$, with respect to fiducial sequences of each GC.
We show our results in Supplementary Figures
\ref{sifig:ocendist} -- \ref{sifig:n6397dist}.
In the figures, the blue horizontal bars denote the mean measurement errors
including propagation errors during the photometric calibrations
with a 2$\sigma_*$ range ($\pm$ 1$\sigma_*$) for individual stars
at given magnitude bins.

In Supplementary Table~\ref{tab:error}, we show comparisons of
the observed FWHMs of RGB stars and
the measurement errors at the magnitude level of horizontal branch stars,
$V_{HB} + 0.5$ $\geq$ $V$  $\geq$ $V_{HB} - 0.5$, in each GC.
To calculate the FWHMs of RGB stars in GCs,
we used the following relation,
\begin{equation}
{\rm FWHM(RGB)} \approx 2.3548\times\sigma_\Delta,
\end{equation}
where $\sigma_\Delta$ is the standard deviation of RGB stars
in the $\Delta(b-y)$ or the $\Delta hk$ distributions in a fixed magnitude bin.
Note that the FWHM of RGB stars in the $hk$ index is slightly larger than
the $hk$ separation between the double populations such as in
M22, NGC1851, NGC2808, M4 and NGC6752.
The mean measurement errors, $\sigma_*(b-y)$ or $\sigma_*(hk)$, given in the table
include propagation errors during the photometric calibrations
and they are those for individual stars.
Therefore, the mean measurement errors for an individual population in
GCs, $\sigma_p(b-y)$ or $\sigma_p(hk)$, will be given by 
$\approx \sigma_*(b-y)$/$\sqrt{n_p}$ or $\sigma_*(hk)$/$\sqrt{n_p}$,
where $n_p$ ($\geq$ 20) is the number of stars in each population.
Note that, while the FWHMs of most GCs have much larger values 
[$\geq$ 8$\sigma_*(hk)$]
than the measurement errors for individual stars in the $hk$ index,
the FWHM of NGC6397 is comparable in size to the measurement error in
the $hk$ index, consistent with the idea that NGC6397 is the only normal GC
with a simple stellar population (i.e. coeval and monometallic) in our sample.
Also note that NGC6397 shows similar degree of the full RGB widths
in the $\Delta(b-y)$ and the $\Delta hk$ distributions
(see Supplementary Table \ref{tab:error} and
Supplementary Figure \ref{sifig:n6397dist}).

The last two columns of Supplementary Table~\ref{tab:error},
$E(b-y)_{1/2}$ and $E(hk)_{1/2}$, are for the contributions
due to the continuous differential reddening effect assuming
a 50\% variation in the total interstellar reddening across each GC.
The observed FWHMs of GCs in the $(b-y)$ color and in the $hk$ index
can not be explained simultaneously, similar to what shown in
Supplementray Figure \ref{sifig:reddening}.
Therefore, the continuous differential
reddening effect can be ruled out to explain the differences between
the observed FWHMs(RGB) and the measurement errors in GCs.

\subsection{$\Delta hk$ as a probe of multiple stellar populations in GCs}
For $\omega$ Cen, M22 and NGC1851 (see Figure 3 and
Supplementary Figure \ref{sifig:hkvsfeh}),
when the two subpopulations are defined by our $hk$ index
(or our $\Delta hk$ distribution), we can also see the clear division in
spectroscopic elemental abundances.
Similarly, we will discuss that the split or the spread in
the $\Delta hk$ distributions of RGB stars in other clusters
can provide a powerful method to probe the multiple
stellar populations in GCs.

The calcium abundance is the major factor that determines
the $hk$ index or the $\Delta hk$ distribution of RGB stars in a GC
(see discussion below) and Type II supernovae are responsible for
the calcium enrichment in a GC.
As discussed, however, our results do not imply that Type II supernovae
are solely responsible for the chemical enrichment in GCs.
In an attempt to explain the observed large star-to-star lighter elemental
abundance variations (in particular O and Na) in GCs, chemical pollution by
intermediate mass AGB stars$^{9}$ or
fast rotating massive (FRM) stars$^{20}$ has been widely accepted.
It should be reminded that, however, neither AGB nor FRM scenarios
can explain the chemical enrichment of the $\alpha$- and
$r$-process elements in the second generation of the stars.
It is most likely that all three aforementioned mechanisms 
(and perhaps including Type Ia supernovae)
are required to explain elemental abundance patterns found in GCs.
In addition to the chemical enrichment by Type II supernovae,
which is the main results presented here,
if the second generation of the stars in some of our GCs have experienced
the chemical pollution by intermediate-mass AGB or FRM stars,
the lighter elemental abundances, such as oxygen and sodium,
between the two generatrions of stars must have been different.
Furthermore, the variations in [O/Fe] and [Na/Fe] can be as large as
1 dex in some GCs$^{7}$ and the differences in the oxygen and
sodium abundances are easily detectable compared to those
in the heavy elements, such as calcium and iron.

During the last few years, tremendous amount of effort has been
directed at spectroscopic study of RGB stars in GCs,
in particular, using the multi-object spectrograph mounted at VLT.
Among our eight GCs,
NGC2808\cite{carretta2808}, M4\cite{marinom4} and NGC6752\cite{carretta6752} 
have been studied using moderately high resolution spectra
for more than 100 RGB stars.
In Supplementary Figure \ref{sifig:ona}, we show comparisons of
$\Delta hk$ versus O, Na and Fe distributions of the clusters.
In panel (a), we show the plot of $V-V_{\rm HB}$ versus $\Delta hk$
for NGC2808 RGB stars. In the figure, the plus signs denote
the RGB stars with known [O/Fe] and [Na/Fe] ratios\cite{carretta2808}.
From the $\Delta hk$ distribution of RGB stars shown in panel (b),
we define the boundary at $\Delta hk$ = $-$0.05 mag (the vertical dashed line)
assuming that NGC2808 has two major stellar populations as shown in panel (b)
or Supplementary Figure \ref{sifig:n2808dist}.
Similar to the procedure employed in M22 and NGC1851 (see Figure 3),
we define the Ca-w group with smaller $hk$ index and
the Ca-s group with larger $hk$ index and they
are denoted by the blue and the red plus signs, respectively, in panel (a).
In panels (c), (d) and (e), we show the [O/Fe], [Na/Fe] and [Fe/H]
distributions for each group, where the shaded histograms outlined with
blue color are for the Ca-w group and the blank histrograms
outlined with red color are for the Ca-s group.
The Ca-w group has a higher mean oxygen and a lower mean sodium
abundances, while the Ca-s group has a lower mean oxygen and
a higher mean sodium abundances, indicative of the presence of
the proton-capture process at high temperature between the two formation
epochs presumably via intermediate-mass AGB or FRM stars,
where oxygen is depleted by the CNO cycle while sodium
is enriched from the $^{22}$Ne + $^1$H $\rightarrow$ $^{23}$Na reaction.
Our results strongly suggest that they are truly different stellar
populations and not related to, for example, 
our photometric measurement errors and differential reddening effect:
the Ca-w group is the first generation of stars
while the Ca-s group is the second generation of stars
enriched by Type II supernovae (e.g. calcium) and
intermediate-mass AGB or FRM stars (e.g. sodium).
Although the difference in the [Fe/H] distributions between the two groups
does not appear to be as compelling as those in the [O/Fe] and
the [Na/Fe] distributions, the Ca-w group has a slightly lower mean
metallicity than the Ca-s group does.
We performed non-parametric Kolmogorov-Smirnov (K-S) tests
to see if the [Fe/H] distributions of the two populations
in NGC2808 are drawn from the same parent population.
Our calculation shows that the probability of being drawn from
identical stellar populations is 5.5\% for NGC2808,
suggesting that they have different parent populations.

The same results can be found in M4 and NGC6752.
From the comparisons of the [O/Fe] and the [Na/Fe] distributions
between the Ca-w and the Ca-s groups,
it can be seen that the Ca-w groups are the first generations of stars
while the Ca-s groups are the second generations of stars
in the clusters.
We also performed K-S tests for the [Fe/H] distribution of M4,
we obtained that the probability of being drawn from
identical parent populations is 5.5\% for M4,
indicating that each subpopulation in M4
has different parent populations.

\subsection{Recalibration of [Fe/H]$_{hk}$ Based on RGB Stars in $\omega$ Cen
and Metallicity Distributions of Eight GCs}
In our previous study for NGC1851, we showed that the $hk$ index traces
the calcium abundance and, furthermore, it can provide a very powerful method
to distinguish multiple stellar populations in GCs$^{22}$.
However, it can be seen that the full range of $\Delta hk$ increases with
the luminosity of RGB stars (i.e. different temperature or surface gravity),
in particular, in $\omega$ Cen and M22.
Due to the temperature dependency on the $hk$ index versus metallicity relation,
the $\Delta hk$ distributions cannot be directly translated into
the absolute metallicity scale.
Therefore, we calculate the photometric metallicity, [Fe/H]$_{hk}$,
of individual RGB stars in eight GCs using the [Fe/H] relations
on the $hk_0$ versus $(b-y)_0$ plane$^{12, 22}$.

Recently, Johnson \emph{et al.}$^{5}$ studied elemental abundances,
including calcium, of large sample of RGB stars in $\omega$ Cen
using moderately high resolution spectra (R $\approx$ 18,000).
Since $\omega$ Cen contains multiple stellar populations
with very broad metallicity range, $\Delta$[Fe/H] $\approx$ 1.5 dex,
comparisons of our results of RGB stars in $\omega$ Cen with
those of Johnson \emph{et al.} may provide an wonderful opportunity to assess
our photometric metallicity scale using the $hk$ index, [Fe/H]$_{hk}$.
In Supplementary Figure \ref{sifig:hkvsfeh},
we show elemental abundances of 40 RGB stars
in $\omega$ Cen studied by Johnson \emph{et al.} as a function of $\Delta hk$.
As shown in the figure, [Ca/H] and [Fe/H] appear to be well correlated with
$\Delta hk$, indicating that $\Delta hk$ can truly be treated
as the relative calcium abundance or metallicity indicators for RGB stars
with similar luminosities in a GC.
We also show plots of [Fe/H]$_{\rm spec}$ versus [Fe/H]$_{hk}$ and
[Ca/H]$_{\rm spec}$ versus [Fe/H]$_{hk}$ for 32 RGB stars with
sufficiently high signal-to-noise ratios ($\geq$ 100).
We derive linear fits to each relation and we find
\begin{equation}
\rm{[Fe/H]}_{\rm spec} = 0.533\rm{[Fe/H]}_{hk} - 0.775~~~~~~(\sigma = 0.087 \rm{dex}),
\label{eqn:metallicity}
\end{equation}
and
\begin{equation}
\rm{[Ca/H]}_{\rm spec} = 0.587\rm{[Fe/H]}_{hk} - 0.403~~~~~~(\sigma = 0.106 \rm{dex}).
\end{equation}

We recalibrate our photometric metallicity using the equation (\ref{eqn:metallicity}),
[Fe/H]$_{hk,corr}$, and we derive metallicity distribution functions (MDFs)
for eight GCs.
During our calculations of MDFs, we use RGB stars with 
$-$2.0 $\leq$ $V$ $-$ $V_{\rm HB}$ $\leq$ $-$0.5 mag in order to minimize
contamination from off-cluster field and red-clump populations.
We show our results in Supplementary Figure \ref{sifig:hkvsfeh}.
As expected from the $\Delta hk$ distributions, the signs of multiple stellar
populations persist in our MDFs for most GCs.

Finally, cautions are advisable on our MDFs of GCs.
Our metallicity scale is not on the traditional Zinn \& West$^{14}$ scale,
therefore our MDFs for GCs can be different from those from
other photometric or spectroscopic studies.
RGB stars in $\omega$ Cen of Johnson \emph{et al.} have different
individual elemental abundances, which were not taken into consideration
in our [Fe/H]$_{\rm spec}$ versus [Fe/H]$_{hk}$ or
[Ca/H]$_{\rm spec}$ versus [Fe/H]$_{hk}$ relations.
Furthermore, each GCs may have slightly different elemental abundance
ratios and our calibrated photometric metallicities for GCs would be affected.
However, it should be emphasized that the crux of our results
is the split or the spread in the $hk$ index in the RGB stars of individual GCs,
which is insensitive to other elemental abundances except
calcium as will be discussed below.

\section{The Influence of Elemental Abundances on the \emph{hk} Index}
The realistic modeling of the resonance Ca II H \& K lines requires
proper understanding of stellar atmospheres, including chromospheres,
and hydrodynamic non-local thermodynamic equilibrium treatments,
which have posed difficult problems for decades\cite{linsky}.
Here, we demonstrate that the calcium abundance is the major
factor that determines the $hk$ index of RGB stars using 
1-dimensional plane-parallel stellar atmospheres\cite{castelli}.

\subsection{Calcium}
Using the model atmosphere for the RGB star at the magnitude
level of the horizontal branch with $T_{\rm eff}$ = 4750 K,
$\log g$ = 2.0 (in cgs unit), $v_{\rm turb}$ = 2.0 km/s, [Fe/H] = $-$1.6,
we calculate synthetic spectra for [Ca/Fe] = 0.25, 0.30, 0.35, 0.40, 0.45
and we show some of our synthetic spectra in Supplementary Figure \ref{sifig:al}.
We convolve the filter transmission functions with synthetic spectra and
we obtain the calcium abundance sensitivity on the $hk$ index,
$\partial (hk)$/$\partial$[Ca/H] $\approx$ 0.422 mag/dex.
Note that this result is based on the fixed model
parameters, such as $T_{\rm eff}$, $\log g$, $v_{\rm turb}$, and [Fe/H],
except calcium abundance.
To interpret observed $\Delta hk$ between the two stellar populations in a GC
in terms of different calcium abundances,
proper atmospheric parameters should be taken into consideration.
For example, the two stellar groups with different [Fe/H] and [$s$-process/Fe]
ratios in M22 by Marino \emph{et al.}$^{6}$ have slightly different
elemental abundances and temperatures. The stars in the metal-poor group by
Marino \emph{et al.} have $\langle$[Fe/H]$\rangle$  = $-$1.82,
$\langle$[Ca/Fe]$\rangle$  = +0.25 and
those in the metal-rich group have $\langle$[Fe/H]$\rangle$  = $-$1.68,
$\langle$[Ca/Fe]$\rangle$  = +0.35.
The mean temperature of the stars in the metal-poor group is
$\sim$ 100 $\pm$ 42 K hotter than those in the metal-rich group.
Using these atmospheric parameters,
we obtain $\Delta hk$ = 0.121 $\pm$ 0.072 mag, which is apparently
consistent with the double peaks in the $\Delta hk$ distribution
of M22 within the error as shown in Supplementary Figure \ref{sifig:m22dist}.

\subsection{Helium}
As shown in Supplementary Figure \ref{sifig:iso},
helium is very important in stellar structure and evolution.
Very unfortunately, however, there is
no direct method to measure helium abundances of stars in GCs.
As we discussed, all the GCs with signs of multiple
stellar populations have relatively EHB, for example,
the second generation of the stars in M22 is expected to have
enhanced helium abundance by $\Delta Y$ $\approx$ 0.05 inferred
from its EHB morphology.
Using the model atmospheres with enhanced helium abundance
($Y$ $\approx$ 0.35, equivalent to $\Delta Y$ $\approx$ 0.10)
by Castelli\footnote{http://wwwuser.oat.ts.astro.it/castelli/grids.html},
we obtain $\Delta hk$ $\approx$ $-$0.002 mag, in the sense that the $hk$ index
decreases as helium abundance increases, and thus
the effect of enhanced helium abundance on the $hk$ index is negligible.
Given the cool temperatures of RGB stars in GCs,
the helium enhancement by $\Delta Y$ = 0.05 -- 0.10
does not appear to be important.

\subsection{CNO}
It is well-known fact that many GCs show large star-to-star
elemental abundance variations. In particular, almost all GCs
show variations in the CNO abundances resulted from the internal 
evolutionary mixing accompanied with the CNO-cycle
or the primordial pollution by intermediate-mass AGB stars
to the second generation of the stars\cite{kraft94, yong}.
In spite of their high abundances, the CNO abundances are hard to measure
in the optical wavelength mainly due to the lack of atomic transitions.
On the other hand, in the form of molecules,
the CNO can affect the $hk$ index,
in particular the CN band at $\lambda$ 3885 \AA~
as shown in Supplementary Figure \ref{sifig:filter}.
Both carbon and nitrogen contribute in the formation of CN molecules.
The typical RGB stars in GCs show
an anticorrelation between the CN band and the CH band
strengths and a correlation between the CN band and the NH
band strengths, indicating that the nitrogen controls the CN
band strength\cite{briley}.
Our results suggest that the variations in the CNO abundances
do not affect the $hk$ index significantly.
We obtain $\Delta hk$ $\approx$ $-$0.007, +0.002, $-$0.004 for
$\Delta$[C,N,O/Fe] = +1.0 dex, respectively,
and their influence on the $hk$ index appears to be negligible.

\subsection{Aluminium}
It is also well-known fact that many GCs show large star-to-star
aluminium variations by more than $\Delta$[Al/Fe] $\approx$ 1.0 dex,
presumably resulted from the proton-capture process
at high temperature
or the primordial pollution by intermediate-mass AGB stars
to the second generation of the stars\cite{kraft97, yong}.
The resonance lines of Al I at $\lambda$ 3944.01 and 3961.52 \AA~ are
often very strong (see Supplementary Figure \ref{sifig:al}) and
it may affect our conclusions that the $hk$ index
traces calcium abundances of RGB stars in GCs.
We obtain $\Delta hk$ $\approx$ 0.013 mag for $\Delta$[Al/Fe] = +1.0 dex.
The effect of the variations in aluminium abundances
on the $hk$ index is insignificant compared to our observations,
by more than a factor of ten.
The insignificant influence of aluminium on the Ca II H \& K lines 
was also confirmed by others for the globular cluster NGC6752$^{17}$.

\subsection{{\bf$\alpha$}-elements}
The $\alpha$-elements (O, Ne, Mg, Si, S, Ar, Ca, and Ti)
are quite abundant and they are major donors to the H$^-$ opacity
in RGB stars. However, their influence, except for Ca, on the $hk$ index
appears to be small.
We obtained $\Delta hk$ $\approx$ $-$0.008 mag for
+0.3 dex variation in $\alpha$-elements excluding calcium.

\subsection{\emph{s}-process elements}
RGB stars in GCs show large star-to-star $s$-process elemental abundance
variations presumably resulted from the primordial pollution by intermediate-mass
AGB stars to the second generation of the stars$^{9, }$\cite{yong}.
For example, globular clusters M22 and NGC1851 show bimodal
$s$-process elemental abundance distributions with
$\Delta$[$s$-process/Fe] $\geq$ 0.5 dex.
We obtained $\Delta hk$ $\approx$ +0.008 mag for
$\Delta$[$s$-process/Fe] = +0.5 dex, and thus their influence
on the $hk$ index is small.

\newcommand{\nodata}{{......}}
\renewcommand{\tablename}{{\bf Supplementary Table}}
\spacing{1.5}

\clearpage
\begin{table}
\begin{center}
{\scriptsize
\begin{tabular}{lcc}
\hline
\hline
\multicolumn{1}{c}{} & \multicolumn{1}{c}{Uncertainty on the $hk$ index} &
\multicolumn{1}{c}{Note}\\
\hline
Photometry & $\leq$ 0.020 mag & random \\
Shifted CTIO $Ca$ passband & $<$ 0.011 mag & systematic \\
Mean radial motion of GCs   & $<$ 0.006 mag & systematic \\
Internal velocity dispersion & $<$ 0.001 mag & random \\
Interstellar Ca II absorption & $<$ 0.010 mag & random \\
\hline  
total  & $\leq$ 0.024 mag & \\
total (random)  & $\leq$ 0.022 mag & \\
\hline  
\hline
\end{tabular}
\caption{Summary of uncertainties relevant to the $hk$ index.
\label{tab:exterr}}}
\end{center}
\end{table}

\clearpage
\begin{table}
\begin{center}
{\scriptsize
\begin{tabular}{lccrrrrrrrrl}
\hline
\hline
\multicolumn{1}{c}{ID} & \multicolumn{1}{c}{$V_{HB}$} &
\multicolumn{1}{c}{$E(B-V)$} &
\multicolumn{5}{c}{Exposure Time (sec)} &
\multicolumn{1}{c}{} &
\multicolumn{2}{c}{Obs. Pos.} &
\multicolumn{1}{c}{Date}\\
\cline{4-8}\cline{10-11}
\multicolumn{1}{c}{}     & \multicolumn{1}{c}{}     &
\multicolumn{1}{c}{}     & \multicolumn{1}{c}{$Ca$} &
\multicolumn{1}{c}{$u$}  & \multicolumn{1}{c}{$v$}  &
\multicolumn{1}{c}{$b$}  & \multicolumn{1}{c}{$y$}  &
\multicolumn{1}{c}{} &
\multicolumn{1}{c}{RA} &
\multicolumn{1}{c}{DEC} &
\multicolumn{1}{c}{(MM/YY)} \\
\hline
$\omega$ Cen & 14.53 & 0.12 & 12,860 & \nodata & 4,890 & 2,130 & 1,300 && 13:26:44 & $-$47:26:28 & 05/07, 02/08 \\
M22          & 14.15 & 0.34 &  8,100 &   2,400 & 1,200 & 2,530 & 1,500 && 18:36:29 & $-$23:55:34 & 07/08, 08/08 \\
NGC1851      & 16.09 & 0.02 & 19,100 &  12,300 & 7,400 & 7,100 & 3,810 &&  5:14:14 & $-$40:01:49 & 02/08, 08/08 \\
NGC2808      & 16.22 & 0.22 & 10,820 & \nodata & 3,600 & 4,960 & 3,080 &&  9:11:57 & $-$64:49:24 & 05/07 \\
M4           & 13.45 & 0.36 &  8,400 &   5,400 & 5,570 & 2,920 & 2,060 && 16:23:33 & $-$26:30:47 & 05/07, 08/08 \\
M5           & 15.07 & 0.03 &  9,660 &   3,900 & 2,100 & 4,010 & 2,390 && 15:18:29 &     2:04:03 & 05/07, 08/08 \\
NGC6752      & 13.70 & 0.04 &  7,500 &   2,400 & 1,800 & 2,400 & 1,200 && 19:10:57 & $-$60:00:20 & 07/08, 08/08 \\
NGC6397      & 12.87 & 0.18 &  3,560 &   3,560 & 2,140 & 1,355 &   930 && 17:40:52 & $-$53:36:06 & 08/06, 09/07 \\
\hline  
\hline
\end{tabular}
\caption{Journal of observations for eight GCs.
Only one field has been observed for a particular GC and
the coordinates are given in columns (9) and (10).
\label{tab:log}}}
\end{center}
\end{table}

\clearpage
\begin{table}
\begin{center}
{\scriptsize
\begin{tabular}{lcccccccc}
\hline
\hline
\multicolumn{1}{c}{ID} & \multicolumn{2}{c}{FWHM(RGB)} &
\multicolumn{1}{c}{}   & \multicolumn{2}{c}{Measurement errors} &
\multicolumn{1}{c}{}   & \multicolumn{2}{c}{Differential Reddening}\\
\cline{2-3} \cline{5-6} \cline{8-9}
\multicolumn{1}{c}{} & 
\multicolumn{1}{c}{$(b-y)$} & \multicolumn{1}{c}{$hk$} & 
\multicolumn{1}{c}{} &
\multicolumn{1}{c}{$\sigma_*(b-y)$} & \multicolumn{1}{c}{$\sigma_*(hk)$} &
\multicolumn{1}{c}{} &
\multicolumn{1}{c}{$E(b-y)_{1/2}$} & \multicolumn{1}{c}{$E(hk)_{1/2}$} \\
\hline
$\omega$ Cen & 0.079 & 0.534 && 0.012 & 0.020 &&  0.089 & 0.014 \\
M22          & 0.050 & 0.216 && 0.004 & 0.008 &&  0.252 & 0.041 \\
NGC1851      & 0.035 & 0.182 && 0.006 & 0.013 &&  0.015 & 0.002 \\
NGC2808      & 0.042 & 0.159 && 0.008 & 0.019 &&  0.163 & 0.026 \\
M4           & 0.037 & 0.119 && 0.005 & 0.010 &&  0.266 & 0.043 \\
M5           & 0.025 & 0.105 && 0.006 & 0.013 &&  0.022 & 0.004 \\
NGC6752      & 0.022 & 0.090 && 0.004 & 0.007 &&  0.030 & 0.005 \\
NGC6397      & 0.024 & 0.034 && 0.007 & 0.012 &&  0.133 & 0.022 \\
\hline
\hline
\end{tabular}
\caption{Comparisons of the observed FWHMs of RGB stars and
the measurement errors at the magnitude level of horizontal branch stars,
$V_{HB} + 0.5$ $\geq$ $V$  $\geq$ $V_{HB} - 0.5$.
The measurement errors are those for individual stars and, therefore,
measurement errors for individual subpopulation in GCs
will be given by $\approx \sigma_*(b-y)$/$\sqrt{n_p}$ or $\sigma_*(hk)$/$\sqrt{n_p}$,
where $n_p$ ($\geq$ 20) is the number of stars in each subpopulation in GCs.
Note that, while the FWHMs of most GCs have much larger values 
[$\geq$ 8$\sigma_*(hk)$]
than the measurement errors for individual stars in the $hk$ index,
the FWHM of NGC6397 is comparable in size to the measurement error in
the $hk$ index, consistent with the idea that NGC6397 is the only normal GC
in our sample (see Supplementary Figures~\ref{sifig:ocendist} --\ref{sifig:n6397dist}).
The last two columns, $E(b-y)_{1/2}$ and $E(hk)_{1/2}$, denote
contributions due to the differential reddening effect assuming
a 50\% variation in the total interstellar reddening of each GC,
with which observed FWHMs of GCs in the $(b-y)$ color and in the $hk$ index
can not be explained simultaneously.
\label{tab:error}}}
\end{center}
\end{table}

\renewcommand{\figurename}{{\bf Supplementary Figure}}
\setcounter{figure}{0}
\spacing{1}

\clearpage
\begin{figure*}
\begin{center}
\includegraphics[scale=.7]{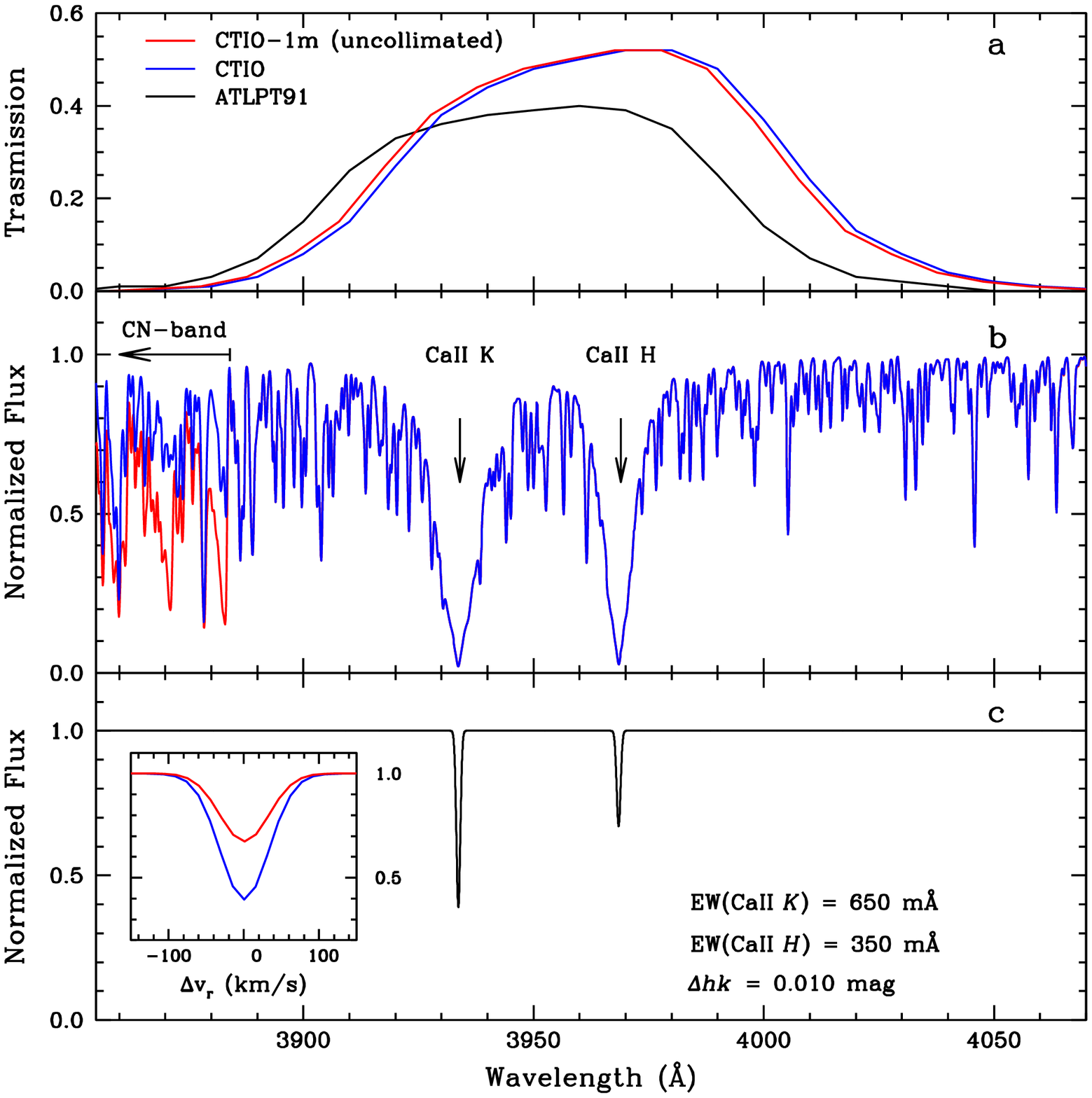}
\end{center}
\caption{(a) A comparison of $Ca$ filter transmission functions between
that in Anthony-Twarog \emph{et al.}$^{12}$ (the black line) and
that for the CTIO 1-m telescope (the blue line).
Both filters have similar FWHMs, approximately 90 \AA,
but the passband for the CTIO 1-m telescope is shifted approximately
15 \AA~ to the longer wavelength.
(b) Synthetic spectra for the CN normal (the blue line) and the CN strong
(the red line) RGB stars.
The CN band at $\lambda$ 3885 \AA~ lies on the lower tail of the $Ca$ filter but
the contamination from the CN band is insignificant.
(c) Synthetic spectra for the interstellar Ca II $H$ \& $K$ lines.
We adopt equivalent widths of 350 m\AA~ and 650 m\AA~ for the interstellar
Ca II $H$ \& $K$ lines, respectively, with a gaussian line profile with
a FWHM of 1 \AA~ (equivalent to $\Delta v_r$ $\approx$ 76 km/s).
In the inset of the figure, the red line denotes the velocity profile
for the interstellar Ca II $H$ line and the blue line for
the interstellar Ca II $K$ line.
This large amount of discrete interstellar absorption
contributes only 0.010 mag to our results.
\label{sifig:filter}}
\end{figure*}

\clearpage
\begin{figure*}
\begin{center}
\includegraphics[scale=.75]{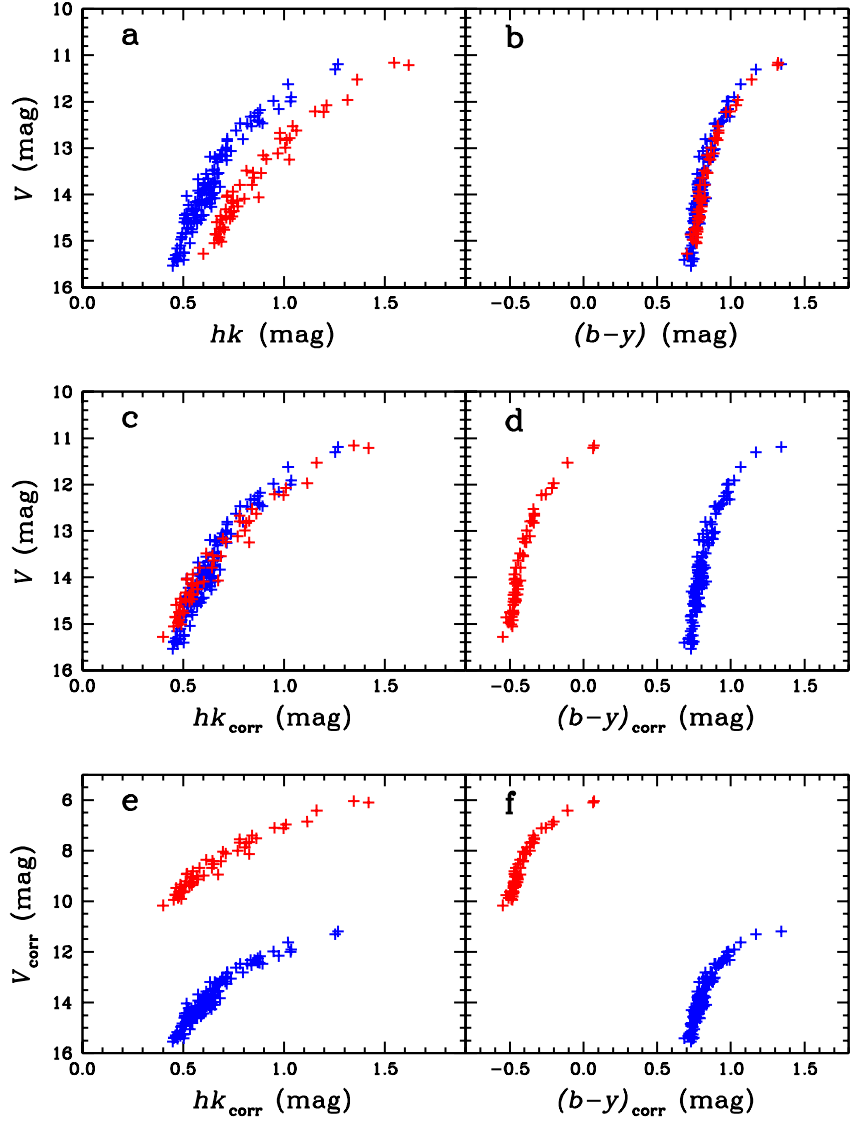}
\end{center}
\caption{(a \& b) Blue crosses and red crosses denote RGB stars 
in the Ca-w and the Ca-s groups, respectively, with proper motion
membership probabilities $P$ $\geq$ 90\%.
(c) RGB stars in the Ca-s group are shifted by $\Delta hk$ = $-$0.20 mag
to match with those in the Ca-w group, assuming the RGB split in M22
is due to differential reddening. The reddening correction value of
$\Delta hk$ = $-$0.20 mag for the Ca-s group is equivalent to
$E(B-V)$ = $-$1.69.
(d) The RGB stars in the Ca-s group are shifted by
$\Delta(b-y)$ = $-$1.25 mag, assuming $E(b-y)/E(B-V)$ = 0.74, and
two RGB sequences do not agree, in the sense that RGB stars
in the Ca-s group is too hot to be in the RGB phase.
(e \& f) After applying reddening correction in $V$ (= $-$5.24 mag),
assuming $A_V$ = 3.1$\times E(B-V)$. The RGB stars in the Ca-s group
become too bright to be members of M22, inconsistent with the proper motion
study of the cluster.
\label{sifig:reddening}}
\end{figure*}

\clearpage
\begin{figure*}
\begin{center}
\includegraphics[scale=.9]{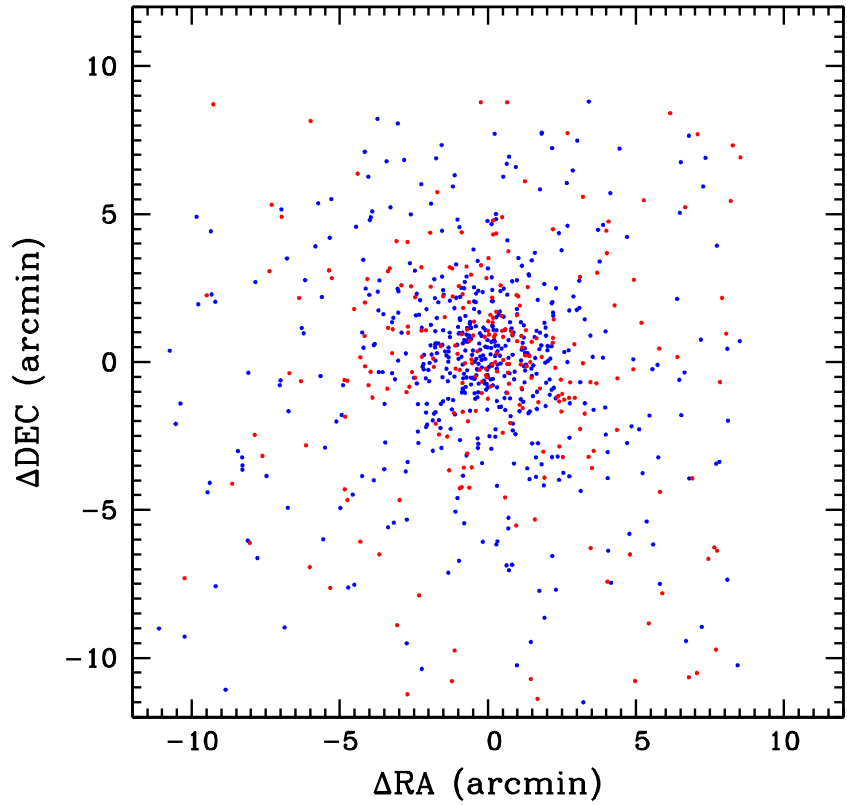}
\end{center}
\caption{Spatial distribution of the Ca-w (blue dots) and the Ca-s
(red dots) groups in M22. Note the absence of spatially patched features,
indicating that differential reddening is not responsible
for the RGB split in M22.
\label{sifig:spatial}}
\end{figure*}

\clearpage
\begin{figure*}
\begin{center}
\includegraphics[scale=.8]{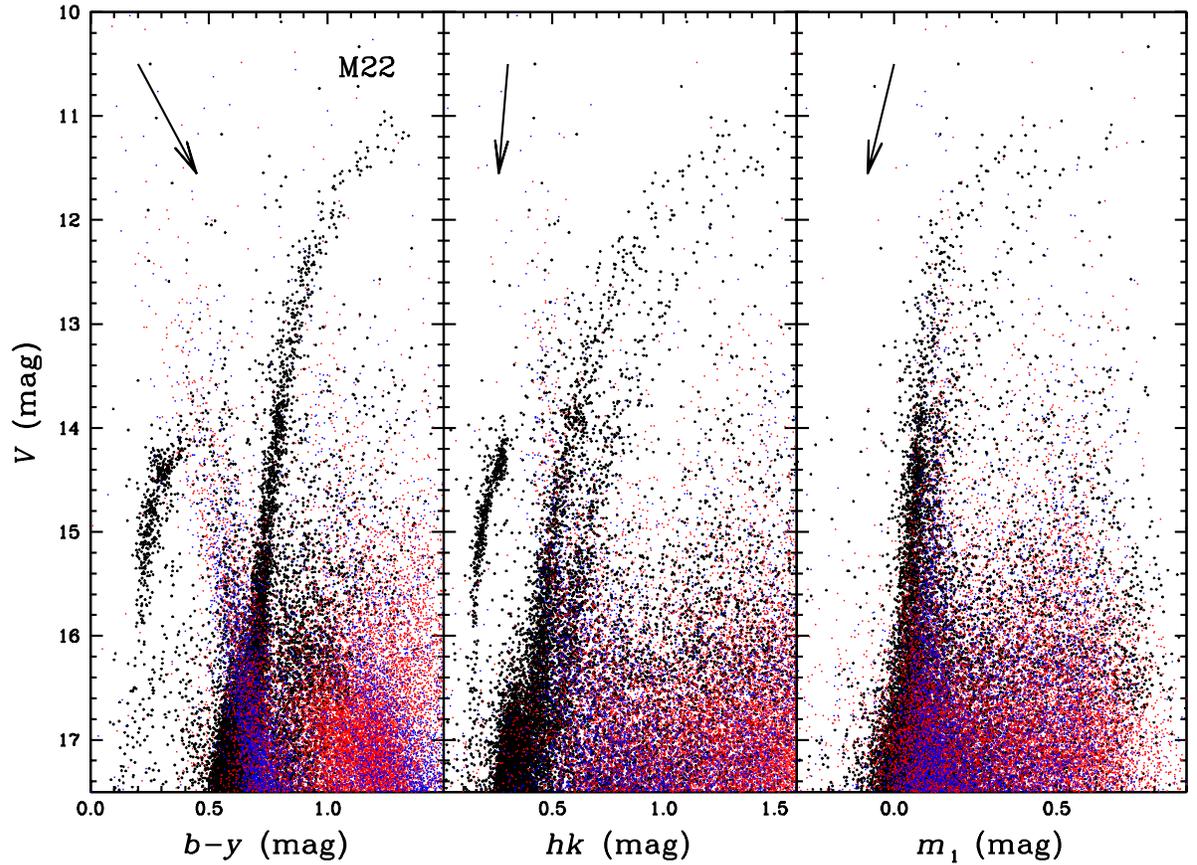}
\end{center}
\caption{Color-magnitude diagrams for M22 and two bulge fields
(NGC6528 and OGLEII-12). The black dots represent M22,
the red dots and the blue dots denote NGC6528 and OGLEII-12, respectively.
The stars in the Milky Way bulge are fainter and redder than those in M22
are and they do not affect the double RGB sequences in M22.
Black arrows indicate reddening vectors, assuming $E(B-V)$ = 0.34
for M22\cite{harris}.
 \label{sifig:m22cmd}}
\end{figure*}

\clearpage
\begin{figure*}
\begin{center}
\includegraphics[scale=.8]{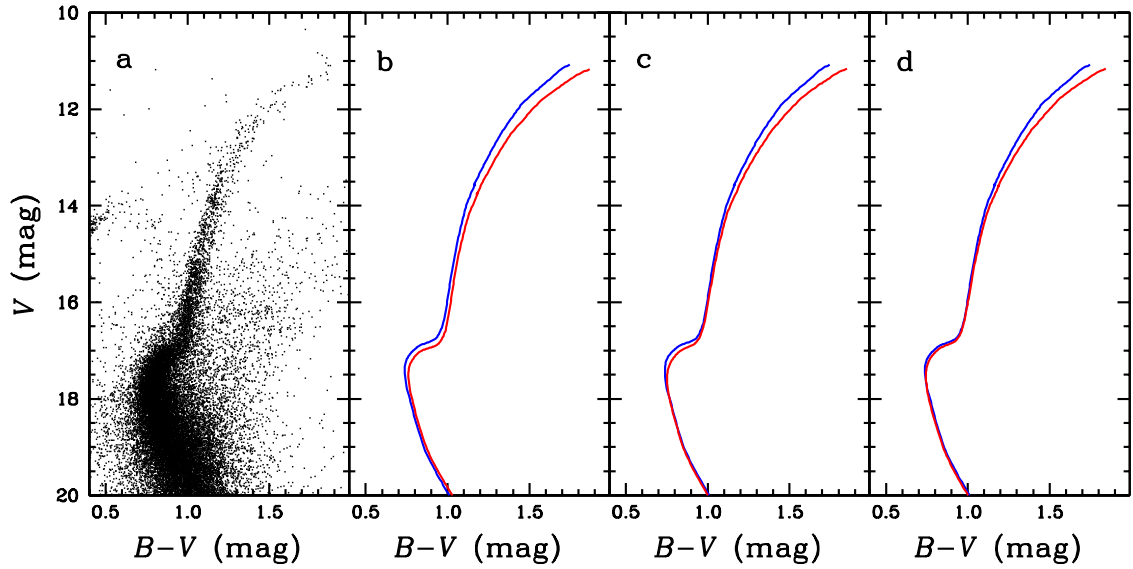}
\end{center}
\caption{(a) The color-magnitude diagram by Monaco \emph{et al.}\cite{monaco}
(b) Model isochrones for [Fe/H] = $-$1.6 (a blue line), and $-$1.4 (a red line),
$Y$ = 0.23, and 11 Gyr.
(c) Model isochrones for [Fe/H] = $-$1.6, $Y$=0.23,
11 Gyr (a blue line) and [Fe/H] = $-$1.4, $Y$=0.28, 11 Gyr (a red line).
(d) Model isochrones for [Fe/H] = $-$1.6, $Y$=0.23, 11 Gyr (a blue line)
and [Fe/H] = $-$1.4, $Y$=0.28, 10 Gyr (a red line).
\label{sifig:iso}}
\end{figure*}

\clearpage
\begin{figure*}
\begin{center}
\includegraphics[scale=.8]{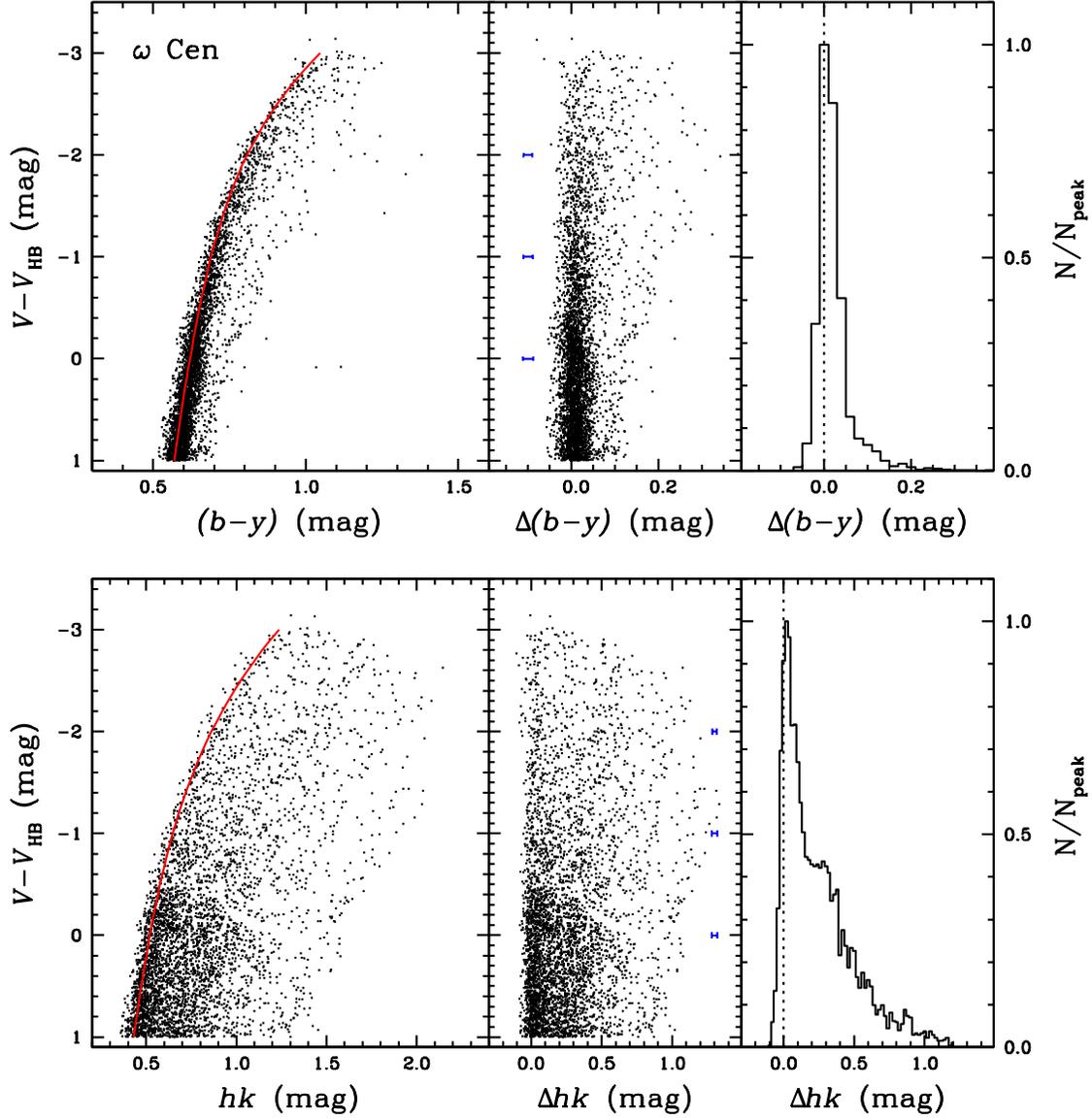}
\end{center}
\caption{The $(b-y)$ and $hk$ distributions of RGB stars in $\omega$ Cen.
The red lines denote fiducial sequences of the cluster, which are forced
to pass through the peak $(b-y)$ colors or $hk$ indices of given magnitude bins.
The $\Delta (b-y)$ and the $\Delta hk$ are differences in the $(b-y)$ color
and the $hk$ index, respectively, of each RGB stars from the fiducial sequences.
The blue horizontal bars indicate measurement errors with a 2$\sigma_*$ range
($\pm$ 1$\sigma_*$) of individual stars at given magnitude bins.
From the $hk$ distribution, at least five distinct populations, whose $hk$
splits are much larger than the measurement errors,
can be found in $\omega$ Cen.
\label{sifig:ocendist}}
\end{figure*}

\clearpage
\begin{figure*}
\begin{center}
\includegraphics[scale=.8]{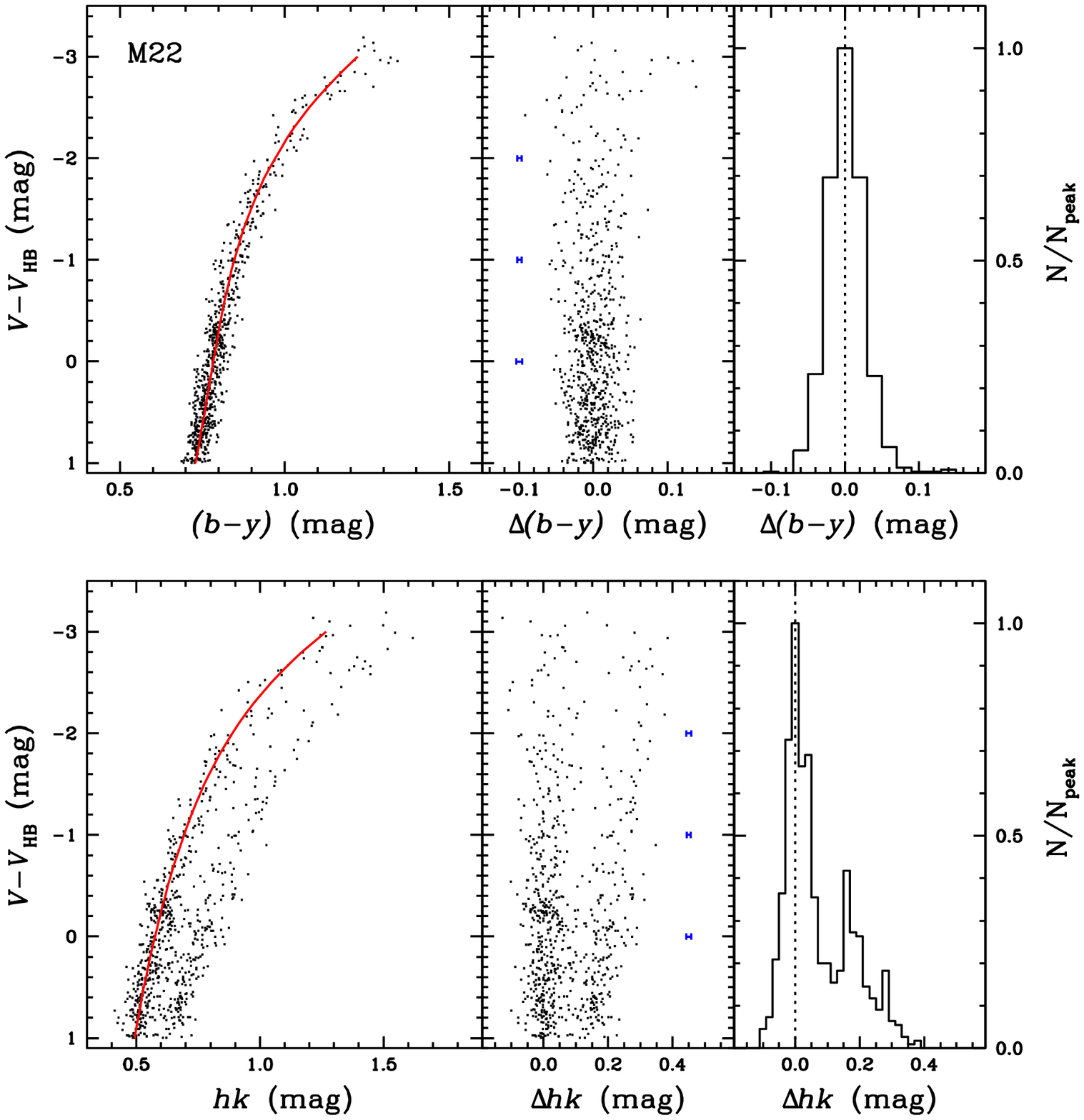}
\end{center}
\caption{The $(b-y)$ and $hk$ distributions of RGB stars in M22.
The blue horizontal bars indicate measurement errors with a 2$\sigma_*$ range
($\pm$ 1$\sigma_*$) of individual stars at given magnitude bins.
Two distinct and discrete populations can be found in M22.
At the magnitude of HB, the $hk$ split between two populations is larger
than 25$\times\sigma_{*}(hk)$ or 250$\times\sigma_{p}(hk)$, where
$\sigma_{*}(hk)$ and $\sigma_{p}(hk)$ denote measurement errors
for individual stars and populations in the $hk$ index, respectively.
\label{sifig:m22dist}}
\end{figure*}

\clearpage
\begin{figure*}
\begin{center}
\includegraphics[scale=.8]{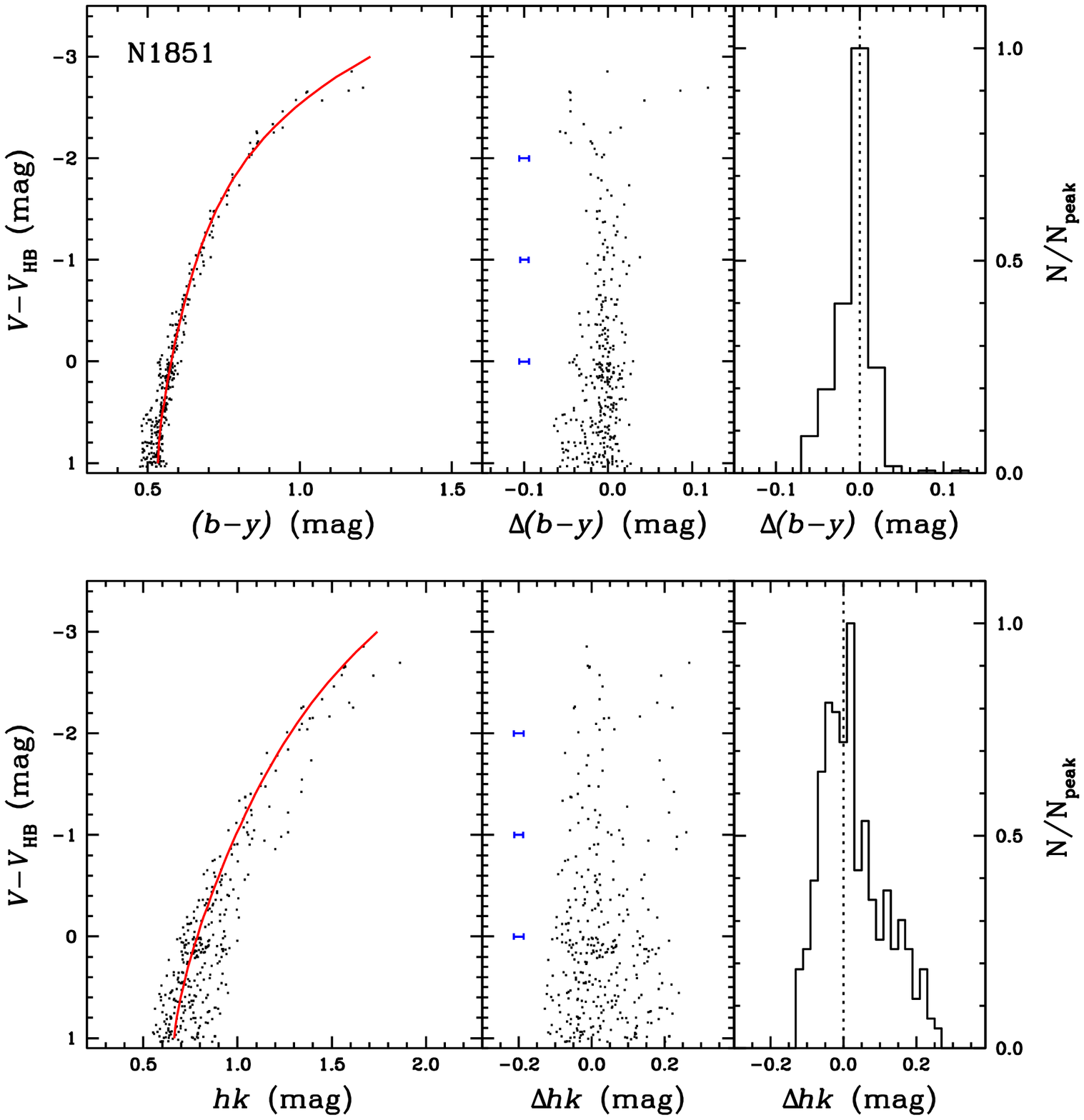}
\end{center}
\caption{The $(b-y)$ and $hk$ distributions of RGB stars in NGC1851.
The blue horizontal bars indicate measurement errors with a 2$\sigma_*$ range
($\pm$ 1$\sigma_*$) of individual stars at given magnitude bins.
Two discrete populations can be found in NGC1851.
At the magnitude of HB, the $hk$ split between two populations is larger
than 11$\times\sigma_{*}(hk)$ or 55$\times\sigma_{p}(hk)$.
\label{sifig:n1851dist}}
\end{figure*}

\clearpage
\begin{figure*}
\begin{center}
\includegraphics[scale=.8]{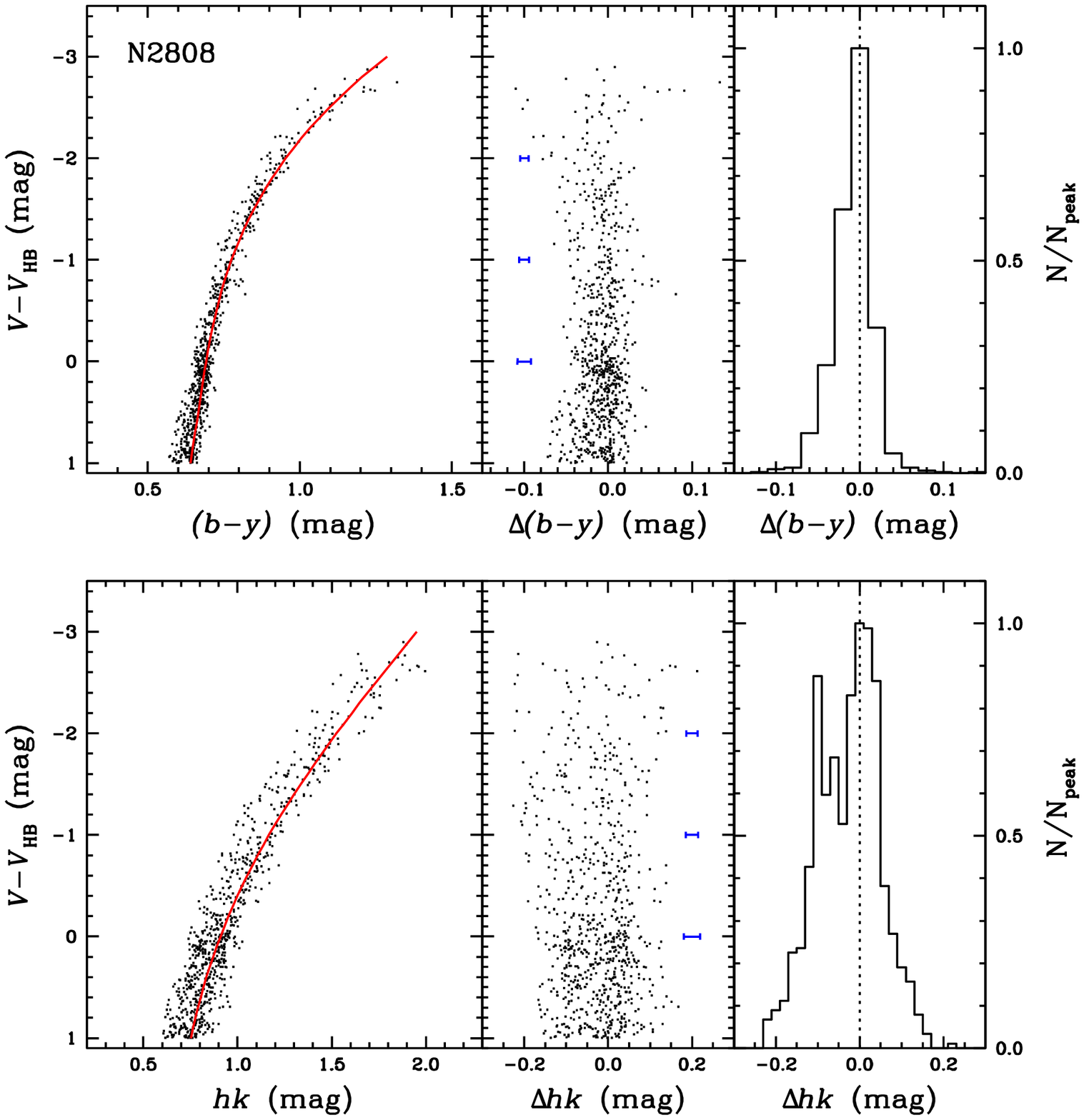}
\end{center}
\caption{The $(b-y)$ and $hk$ distributions of RGB stars in NGC2808.
The blue horizontal bars indicate measurement errors with a 2$\sigma_*$ range
($\pm$ 1$\sigma_*$) of individual stars at given magnitude bins.
At least two discrete populations can be found in NGC2808.
At the magnitude of HB, the $hk$ split between two major populations is larger
than 5$\times\sigma_{*}(hk)$ or 50$\times\sigma_{p}(hk)$.
\label{sifig:n2808dist}}
\end{figure*}

\clearpage
\begin{figure*}
\begin{center}
\includegraphics[scale=.8]{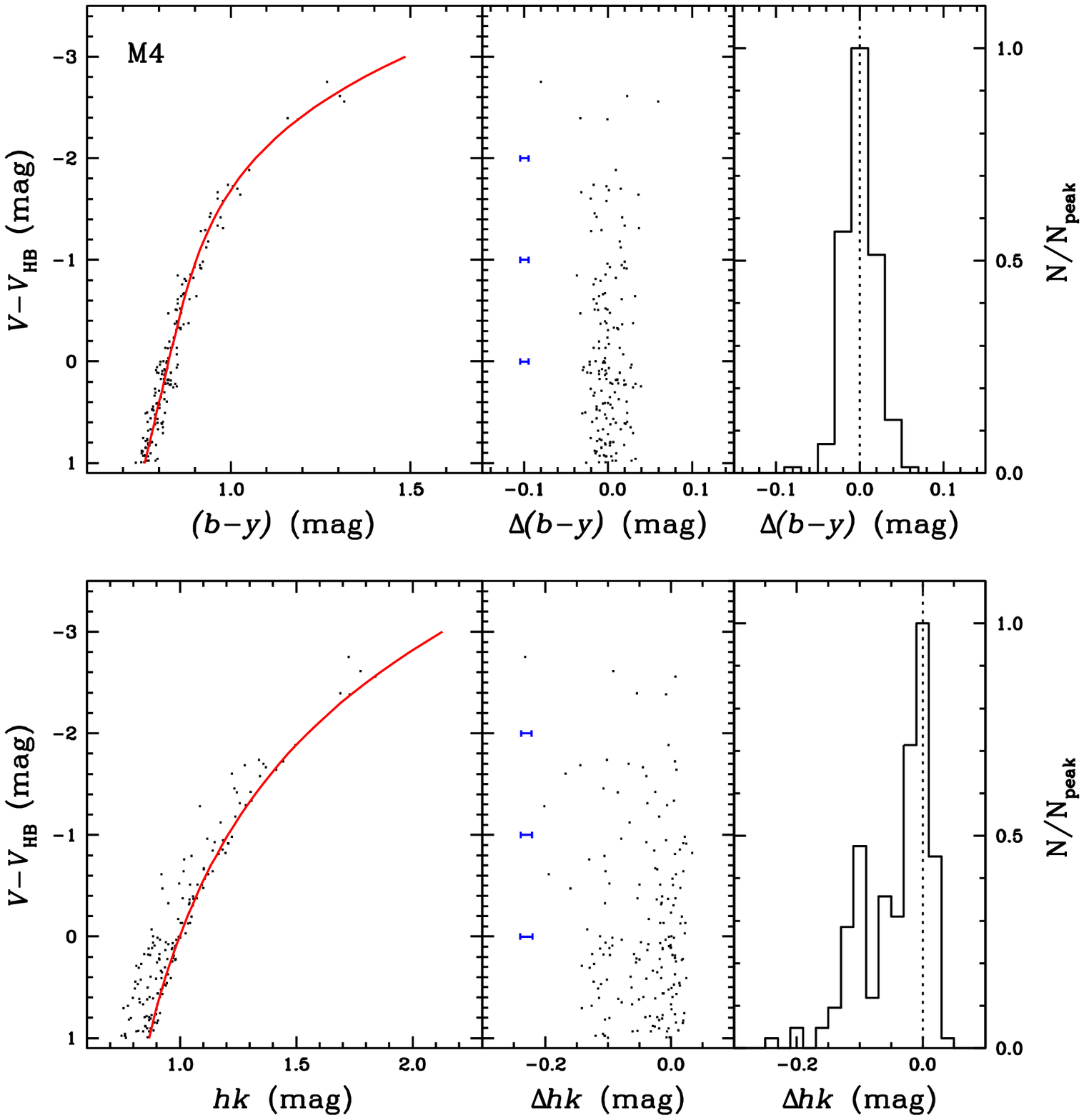}
\end{center}
\caption{The $(b-y)$ and $hk$ distributions of RGB stars in M4.
The blue horizontal bars indicate measurement errors with a 2$\sigma_*$ range
($\pm$ 1$\sigma_*$) of individual stars at given magnitude bins.
Two discrete populations can be found in M4.
At the magnitude of HB, the $hk$ split between two populations is larger
than 10$\times\sigma_{*}(hk)$ or 45$\times\sigma_{p}(hk)$.
\label{sifig:m4dist}}
\end{figure*}

\clearpage
\begin{figure*}
\begin{center}
\includegraphics[scale=.8]{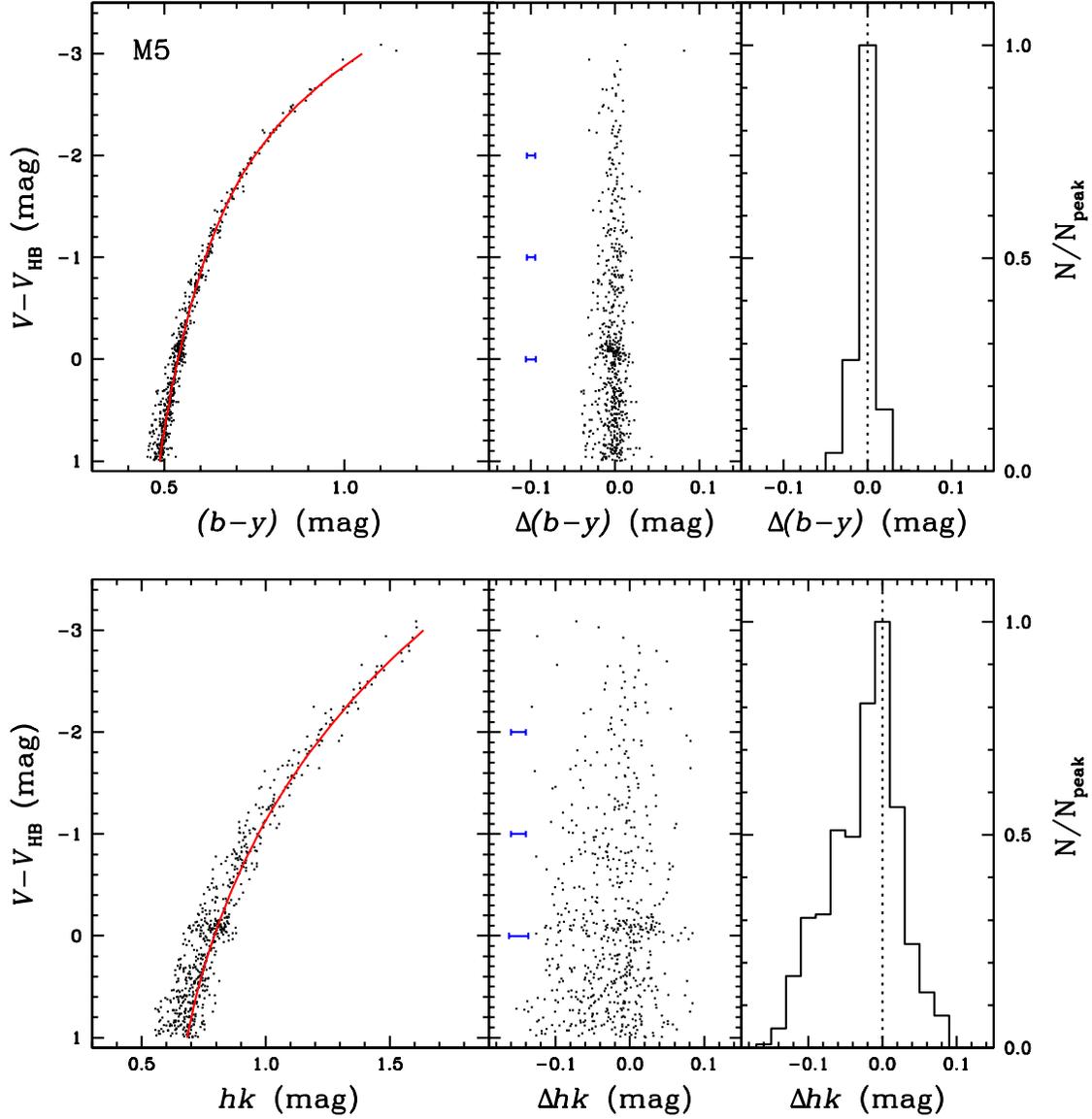}
\end{center}
\caption{The $(b-y)$ and $hk$ distributions of RGB stars in M5.
The blue horizontal bars indicate measurement errors with a 2$\sigma_*$ range
($\pm$ 1$\sigma_*$) of individual stars at given magnitude bins.
The RGB sequence of the cluster shows a large spread in the $hk$ index,
indicative of heterogeneous calcium abundances.
At the magnitude of HB, the FWHM of RGB stars in M5 is larger than
8$\times\sigma_{*}(hk)$.
\label{sifig:m5dist}}
\end{figure*}

\clearpage
\begin{figure*}
\begin{center}
\includegraphics[scale=.8]{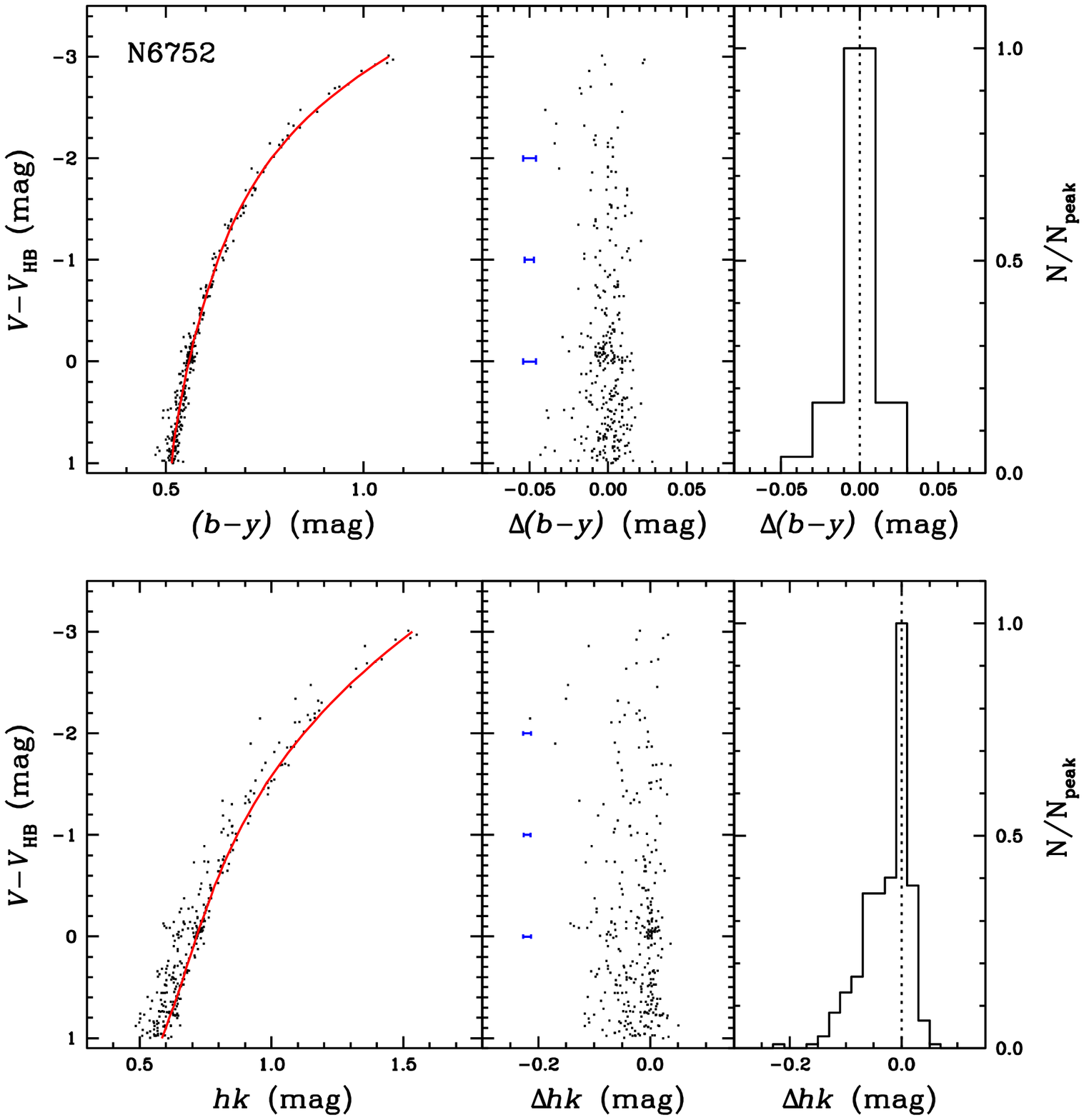}
\end{center}
\caption{The $(b-y)$ and $hk$ distributions of RGB stars in NGC6752.
The blue horizontal bars indicate measurement errors with a 2$\sigma_*$ range
($\pm$ 1$\sigma_*$) of individual stars at given magnitude bins.
Two discrete populations can be found in NGC6752.
At the magnitude of HB, the $hk$ split between two populations is larger
than 10$\times\sigma_{*}(hk)$ or 70$\times\sigma_{p}(hk)$.
\label{sifig:n6752dist}}
\end{figure*}

\clearpage
\begin{figure*}
\begin{center}
\includegraphics[scale=.8]{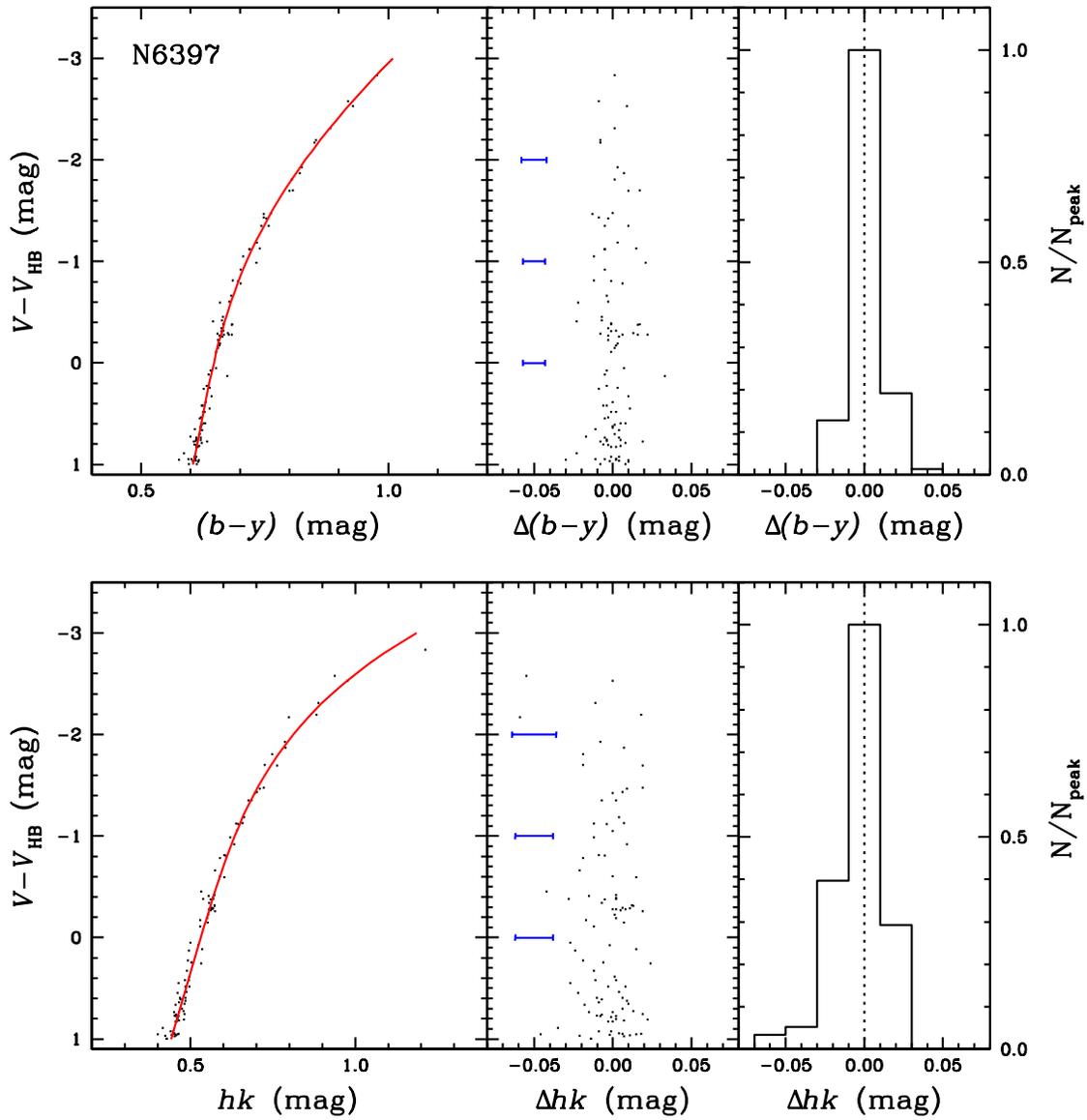}
\end{center}
\caption{The $(b-y)$ and $hk$ distributions of RGB stars in NGC6397.
The blue horizontal bars indicate measurement errors with a 2$\sigma_*$ range
($\pm$ 1$\sigma_*$) of individual stars at given magnitude bins.
It is the only normal GC in Figure 2.
Note the similar degree of the RGB widths in $\Delta(b-y)$ and $\Delta hk$
and the similar degree of RGB widths as the measurement errors.
\label{sifig:n6397dist}}
\end{figure*}

\clearpage
\begin{figure*}
\begin{center}
\includegraphics[scale=.8]{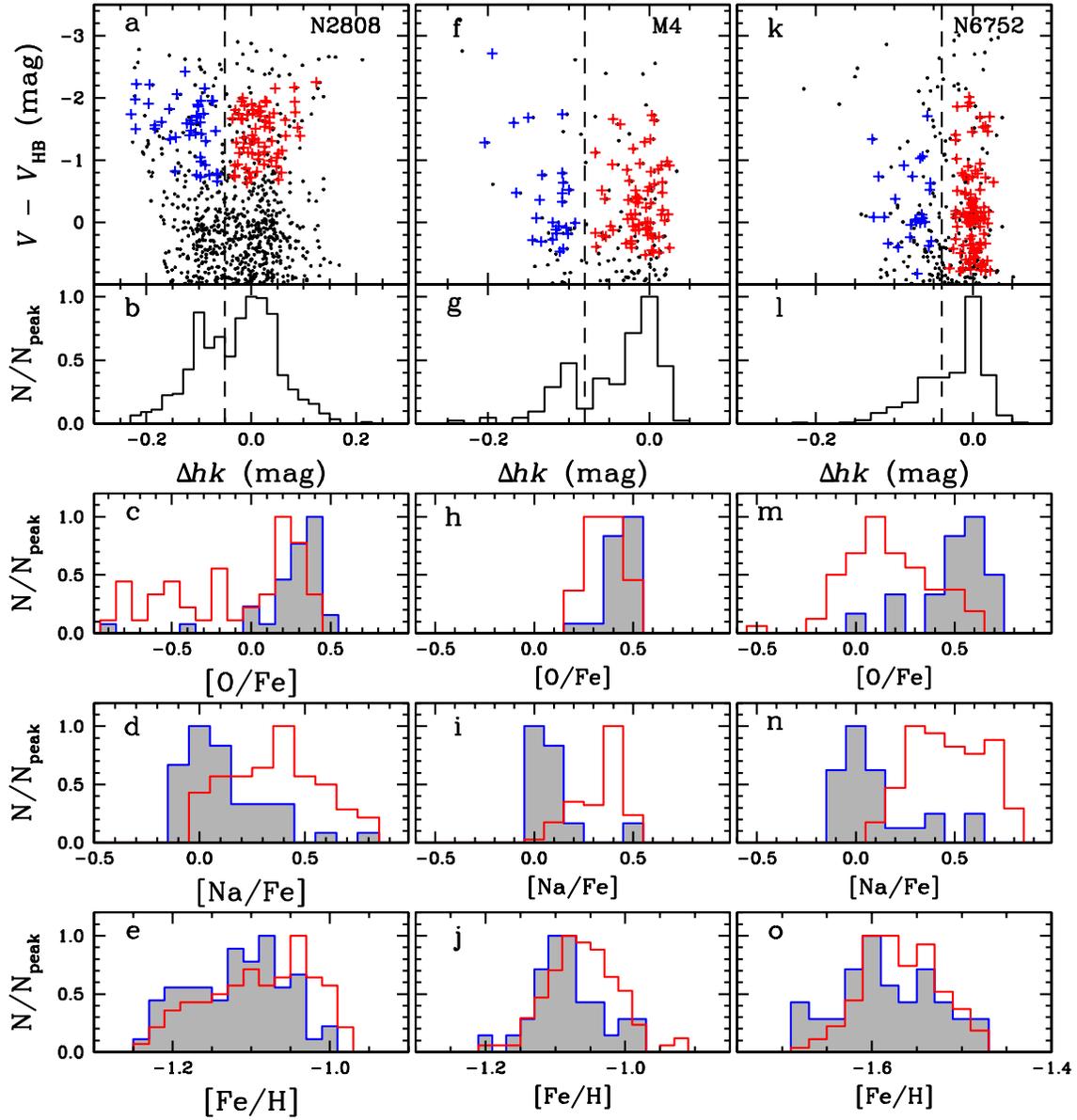}
\end{center}
\caption{(a) A plot of $V-V_{\rm HB}$ versus $\Delta hk$ for RGB stars
in NGC2808. The blue and red plus signs denote the Ca-w and the Ca-s
RGB stars. The dashed line denotes the boundary between the two groups
at $\Delta hk$ = $-$0.05 mag.
(b) The $\Delta hk$ distribution of NGC2808 RGB stars.
(c) The [O/Fe] distributions of the two RGB populations in NGC2808.
The shaded histogram outlined with blue color is for the Ca-w group and
the blank histrogram outlined with red color is for the Ca-s group.
(d) The [Na/Fe] distributions.
(e) The [Fe/H] distributions.
(f--j) Same as (a--e), but for M4 RGB stars 
with the boundary at  $\Delta hk$ = $-$0.08 mag.
(k--o) Same as (a--e), but for NGC6752 RGB stars
with the boundary at  $\Delta hk$ = $-$0.04 mag.
\label{sifig:ona}}
\end{figure*}

\clearpage
\begin{figure*}
\begin{center}
\includegraphics[scale=.8]{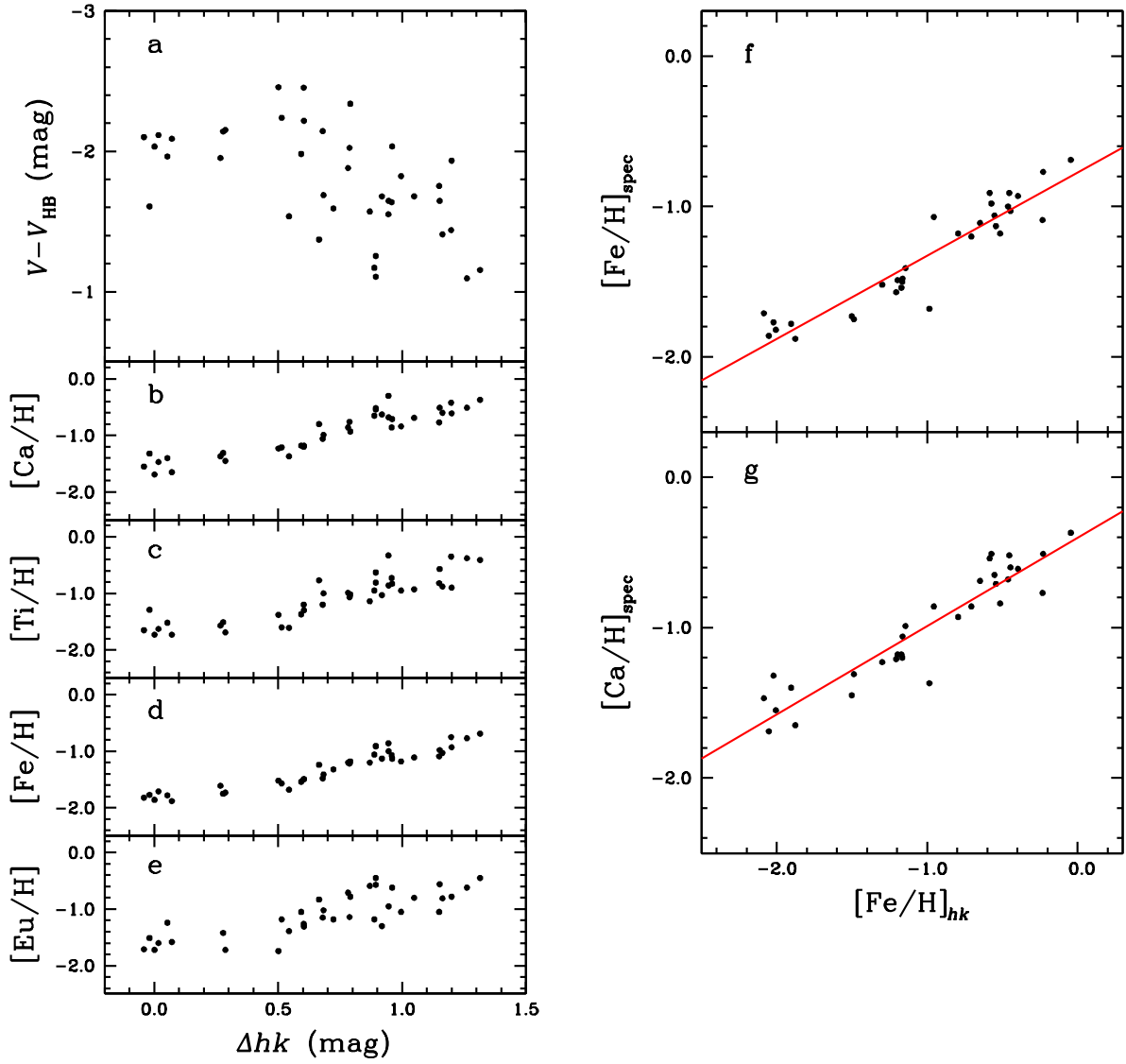}
\end{center}
\caption{(a) A plot of $V-V_{HB}$ versus $\Delta hk$ for 40 RGB stars
in $\omega$ Cen of Johnson et al.$^{5}$
(b -- e) Elemental abundances of 40 RGB stars in $\omega$ Cen
as functions of $\Delta hk$.
(f -- g) Comparisons of our photometric metallicity, [Fe/H]$_{hk}$,
with spectroscopic metallicity, [Fe/H]$_{\rm spec}$, and calcium abundance,
[Ca/H]$_{\rm spec}$. The linear fits to the data are shown with red lines.
\label{sifig:hkvsfeh}}
\end{figure*}

\clearpage
\begin{figure*}
\begin{center}
\includegraphics[scale=.8]{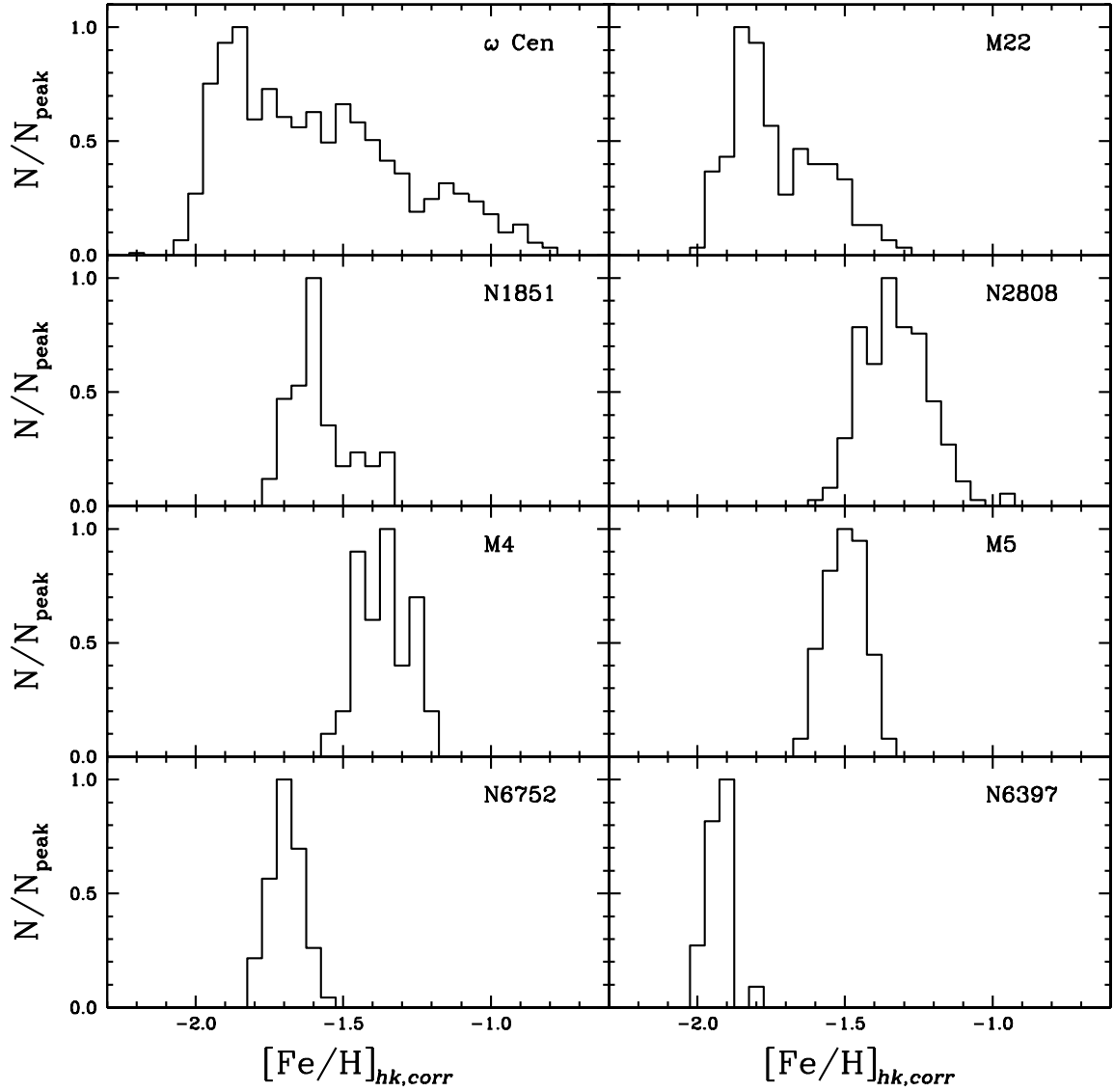}
\end{center}
\caption{Metallicity distribution functions for eight GCs derived from
the $hk$ index. In the figure,
[Fe/H]$_{hk,corr}$ is our recalibrated photometric 
metallicity using the equation (\ref{eqn:metallicity}).
Note that we only use bright RGB stars in order to minimize contamination
from off-cluster field and red-clump populations.
For most GCs, signs of multiple stellar populations persist in our MDFs.
\label{sifig:mdf}}
\end{figure*}

\clearpage
\begin{figure*}
\begin{center}
\includegraphics[scale=.8]{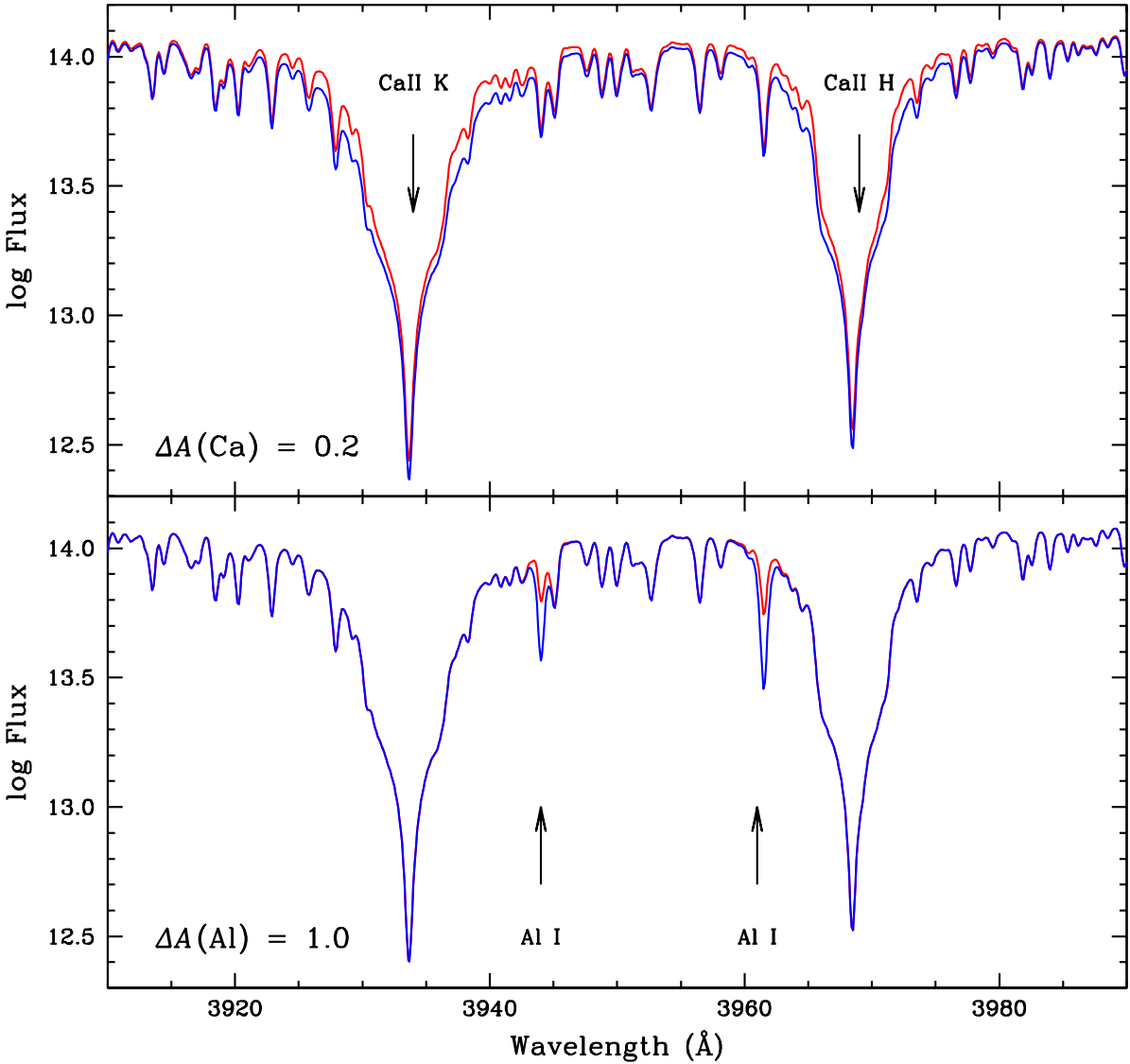}
\end{center}
\caption{Comparisons of synthetic spectra for [Fe/H] = $-$1.6, Teff = 4750 K,
$\log$ g = 2.0.
(Upper panel) The red line denotes synthetic spectrum for [Ca/Fe] = 0.25 dex
and the blue line denotes that for [Ca/Fe] = 0.45 dex.
We adopt the fixed aluminium abundance of [Al/Fe] = 0.50 dex.
(Lower panel) The red line denotes the synthetic spectrum for [Al/Fe] = 0.00 dex
and the blue line denotes that for [Al/Fe] = 1.00 dex.
We adopt the fixed calcium abundance of [Ca/Fe] = 0.30 dex.
The effect of aluminium contamination on the $hk$ index appears to be negligible.
\label{sifig:al}}
\end{figure*}

\end{document}